\documentclass[fleqn,useAMS,usenatbib]{mn2e}
\usepackage{graphicx}
\usepackage{amsmath}
\usepackage{amssymb}

\newcommand{\cs}{c_\mathrm{s}}
\newcommand{\mach}{\mathcal{M}}

\newcommand{\vect}[1]{{\mathbf{#1}}}

\title[Solenoidal and Compressible Modes in Interstellar Clouds]{An Observational Method to Measure the Relative Fractions of Solenoidal and Compressible Modes in Interstellar Clouds}
\author[C. M. Brunt \& C. Federrath]{C. M. Brunt$^{1}$\thanks{E-mail brunt@astro.ex.ac.uk} \& C. Federrath$^{2,3}$ \\
$^{1}$Astrophysics Group, School of Physics, University of Exeter, Stocker Road, Exeter, EX4 4QL, UK\\
$^{2}$Monash Centre for Astrophysics, School of Mathematical Sciences, Monash University, Vic 3800, Australia\\
$^{3}$Zentrum f\"{u}r Astronomie der Universit\"{a}t Heidelberg, Institut f\"{u}r Theoretische Astrophysik, Albert-Uerberle-Str. 2, 69120 Heidelberg, Germany\\
}
\begin{document}

\date{Accepted ; Received ; in original form }

\pagerange{\pageref{firstpage}--\pageref{lastpage}} \pubyear{2011}

\maketitle

\label{firstpage}

\begin{abstract}
We introduce a new method for observationally estimating the fraction of
momentum density (${\rho}{\mathbf{v}}$) power contained in solenoidal modes 
(for which $\nabla \cdot {\rho}{\mathbf{v}} = 0$) in molecular clouds. The method
is successfully tested with numerical simulations of supersonic turbulence that produce
the full range of possible solenoidal/compressible fractions. At present the method assumes statistical
isotropy, and does not account for anisotropies caused by (e.g.) magnetic fields. 
We also introduce a framework for statistically describing density--velocity correlations in 
turbulent clouds. 
\end{abstract}


\section{Introduction}

As the principal sites of star formation in the local universe, molecular
clouds demand much observational and theoretical attention. Their structure is extremely
complex, driven by the interaction of supersonic turbulence, gravity, and magnetic fields
(e.g. Mac Low \& Klessen 2004; Elmegreen \& Scalo 2004; McKee \& Ostriker 2007; Chapman et al 2011; Heyer \& Brunt 2012).
Theoretical descriptions of molecular clouds must begin with the structure of the
relevant physical fields (density, velocity, etc) in three dimensions (3D), while our
practical information is necessarily restricted to what can be measured from the projection
of these fields onto the observational axes -- two spatial and, for spectral line data, 
one (line-of-sight) velocity component. It is a crucially important, yet challenging, 
problem to relate the projected fields to intrinsic properties of the 3D physical fields.

Only limited information about the 3D density field, $\rho(x, y, z)$, can be derived via
analysis of the projected 2D column density field, $N(x, y)$, which is observationally
obtained by extinction measurements, dust emission, or integrated spectral line intensities.
For the latter, we may be aided by (or perhaps limited by) density-selectivity of
particular molecular transitions, yet molecular line emission is the only source of
information for studying the dynamics of molecular clouds.

Arguably the most useful property of $N(x, y)$ is that its Fourier transform, $\tilde{N}(k_{x}, k_{y})$,
is directly proportional to a 2D slice through the Fourier transform of the density field, 
$\tilde{\rho}(k_{x}, k_{y}, k_{z})$, where the line-of-sight wavevector, $k_{z} = 0$;
i.e. $\tilde{N}(k_{x}, k_{y}) \propto \tilde{\rho}(k_{x}, k_{y}, k_{z} = 0)$. This allows,
under the assumption of isotropy, the density power spectrum to be derived (e.g. Stutzki et al 1998), 
the 3D density variance to be inferred (Fischera \& Dopita 2004; 
Brunt, Federrath, \& Price 2010(a); hereafter BFP), and an
estimate of the 3D density PDF to be constructed (Brunt, Federrath, \& Price 2010(b)). 
The essence of the BFP method (relating the 2D normalised column density variance to the
3D normalised density variance) lies in determining the fraction of variance contained in a single
2D slice of the 3D power spectrum.

While BFP focused primarily on a method to relate density and column density statistics, 
they also presented a brief outline of the same method applied to velocity fields, though
noting that the natural observational density-weighting of the velocity field would potentially
cause problems. In this paper, we present an extended development of the outline method
presented in BFP, subject to two key modifications. Firstly, the density-weighting of the velocity
field means that the physical field suitable for BFP-like analysis is the ``momentum
density'' field, ${\rho}{\mathbf{v}}$, rather than the velocity field alone. Secondly,
the realisation that only transverse (solenoidal) modes (for which $\nabla \cdot {\rho}{\mathbf{v}} = 0$)
are projected into 2D allows us to extend the BFP method to estimate the fraction
of momentum density power that is held in solenoidal modes, if given an
estimate of the total momentum density power (through a spectral line imaging observation).
The ideas underpinning the extended method are presented below, along with a demonstration
of its applicability using numerical simulations of turbulent clouds. The method as presented
assumes statistical isotropy, so should not be applied to clouds for which significant anisotropy
is observed or suspected due to (e.g.) the presence of a strong magnetic field at low Mach numbers (BFP).

The layout of this paper is as follows. 
In Section~2, we present the method for observationally
estimating the fraction of momentum power in transverse modes, after first introducing the
background considerations necessary for its formulation. In Section~3, we describe the 
numerical simulations to be used for testing purposes. In Section~4, we test the method by applying it to the
numerical simulations, followed by a discussion (Section 5) and summary (Section 6).
In the Appendix we examine statistical aspects of density--velocity correlation, which
has a small effect on the method.

\section{Solenoidal and Compressible Modes}

In this Section we discuss geometrical properties of divergence--free (``transverse'' or ``solenoidal'') and
curl--free (``longitudinal'' or ``compressible'') modes in a general vector field (Section 2.1), before considering the
consequences of projection of such a field from 3D to 2D (Section 2.2). In the following we will use transverse/solenoidal
and longitudinal/compressible to refer respectively to divergence--free and curl--free components. Subsequently, we identify the 
momentum field as the most relevant physical field of interest for quantitative analysis and develop a method
by which the fraction of power in transverse momentum modes may be estimated observationally (Section 2.3).

\subsection{General Considerations}

The Helmholtz Decomposition Theorem (Helmholtz 1858) states that, as a function of position 
$\mathbf{x} = (x, y, z)$, an arbitrary 3D vector field, 
$\mathbf{F}(\mathbf{x})$, can be represented as the sum of a purely transverse 
field, $\mathbf{F}_{\perp}(\mathbf{x})$, and a purely longitudinal field, $\mathbf{F}_{||}(\mathbf{x})$:
\begin{equation}
\mathbf{F}(\mathbf{x}) = \mathbf{F}_{\perp}(\mathbf{x})  + \mathbf{F}_{||}(\mathbf{x}) , 
\label{helmholtzdec}
\end{equation}
where:
\begin{equation}
\mathbf{\nabla} \cdot \mathbf{F}_{\perp} = 0 ,
\label{dot1}
\end{equation}
\begin{equation}
\mathbf{\nabla} \times \mathbf{F}_{||} = 0 .
\label{cross1}
\end{equation}
In Fourier space, the equivalent relations to equations~(\ref{dot1}) and (\ref{cross1}) 
are (using a tilde to represent the Fourier transformed fields):
\begin{equation}
\tilde{\mathbf{F}}(\mathbf{k})  = \tilde{\mathbf{F}}_{\perp}(\mathbf{k})  + \tilde{\mathbf{F}}_{||}(\mathbf{k}) ,
\end{equation}
\begin{equation}
\mathbf{k} \cdot \tilde{\mathbf{F}}_{\perp} = 0 ,
\label{dot2}
\end{equation}
\begin{equation}
\mathbf{k} \times \tilde{\mathbf{F}}_{||} = 0 ,
\label{cross2}
\end{equation}
where $\mathbf{k} = (k_{x}, k_{y}, k_{z})$ are wavevectors, and the Fourier transformed field
is defined, over a cubical spatial region of length $L$ on each axis, by:
\begin{equation}
\tilde{\mathbf{F}}({\mathbf{k}}) =  \displaystyle\int_{-L/2}^{L/2}\displaystyle\int_{-L/2}^{L/2}\displaystyle\int_{-L/2}^{L/2} \; {\mathrm{d}}^{3}{\mathbf{x}} \; {\mathbf{F}}({\mathbf{x}}) \; {\mathrm{e}}^{ -2{\mathrm{\pi}} {\mathrm{i}} {\mathbf{k}} \cdot  {\mathbf{x}} / L} .
\label{ftf}
\end{equation}

In the following, we will assume that a reference frame can be chosen in which $\langle {\mathbf{F}} \rangle = 0$,
so that the spatial average, $\langle {\mathbf{F}}^{2} \rangle$, is equal to the field variance, $\sigma^{2}_{\mathbf{F}}$.
It is possible, making use of Parseval's Theorem, to calculate the field variances via Fourier space using:
\begin{equation}
\sigma_{{\mathbf{F}}}^{2} = {\frac{1}{L^{6}}} \displaystyle\sum_{k_{x}=-\infty}^{\infty} \sum_{k_{y}=-\infty}^{\infty} \sum_{k_{z}=-\infty}^{\infty} \; \tilde{{\mathbf{F}}} \cdot \tilde{{\mathbf{F}}}^{*}  ,
\label{parseval1}
\end{equation}
\begin{equation}
\sigma_{{\mathbf{F}}_{\perp}}^{2} = {\frac{1}{L^{6}}} \displaystyle\sum_{k_{x}=-\infty}^{\infty} \sum_{k_{y}=-\infty}^{\infty} \sum_{k_{z}=-\infty}^{\infty} \; \tilde{{\mathbf{F}}}_{\perp} \cdot \tilde{{\mathbf{F}}}_{\perp}^{*}  ,
\end{equation}
\begin{equation}
\sigma_{{\mathbf{F}}_{||}}^{2} = {\frac{1}{L^{6}}} \displaystyle\sum_{k_{x}=-\infty}^{\infty} \sum_{k_{y}=-\infty}^{\infty} \sum_{k_{z}=-\infty}^{\infty} \; \tilde{{\mathbf{F}}}_{||} \cdot \tilde{{\mathbf{F}}}_{||}^{*}  .
\label{parseval2}
\end{equation}
Note that the {\it local} orthogonality of $\tilde{\mathbf{F}}_{\perp}$ and $\tilde{\mathbf{F}}_{||}$ (ensuring that 
$\sigma^{2}_{{\mathbf{F}}} = \sigma^{2}_{{\mathbf{F}}_{\perp}} + \sigma^{2}_{{\mathbf{F}}_{||}}$) does not mean that the direct space
fields ${\mathbf{F}}_{\perp}$ and ${\mathbf{F}}_{||}$  are also locally orthogonal. That is,
at any single field point, we have:
\begin{equation}
F^{2} = {\mathbf{F}} \cdot {\mathbf{F}} = F^{2}_{\perp} + F^{2}_{||} + 2 {\mathbf{F}}_{\perp} \cdot {\mathbf{F}}_{||} \neq F^{2}_{\perp} + F^{2}_{||} ,
\label{equation11}
\end{equation}
since in general ${\mathbf{F}}_{\perp} \cdot {\mathbf{F}}_{||} \neq 0$ {\it locally}. However, this dot product
vanishes when averaged over the entire space containing ${\mathbf{F}}$. (This is required by Parseval's Theorem,
expressed in equations~(\ref{parseval1}--\ref{parseval2}). Note that $\tilde{{\mathbf{F}}} \cdot \tilde{{\mathbf{F}}}^{*} = 
\tilde{{\mathbf{F}}}_{\perp} \cdot \tilde{{\mathbf{F}}}_{\perp}^{*} + \tilde{{\mathbf{F}}}_{||} \cdot \tilde{{\mathbf{F}}}_{||}^{*}$ 
is satisfied at each point in Fourier space, so that $\sigma_{{\mathbf{F}}}^{2} = \langle{\mathbf{F}} \cdot {\mathbf{F}} \rangle = 
\sigma_{{\mathbf{F}}_{\perp}}^{2} + \sigma_{{\mathbf{F}}_{||}}^{2} = \langle F^{2}_{\perp} \rangle + \langle F^{2}_{||} \rangle$,
provided the averages are computed in the $\langle {\mathbf{F}} \rangle = 0$ frame, as we have assumed.)

The Helmholtz Decomposition (equation~(\ref{helmholtzdec})) is unique, up to a vector constant, provided that the
field $\mathbf{F}$ falls to zero on its outer boundary. More generally, $\mathbf{F}$ could
in principle contain a contribution from a curl--free, divergence--free component, $\mathbf{F}_{L}$, given by
the gradient of a scalar harmonic field $\phi_{L}$ that satisfies the Laplace equation (${\nabla}^{2}\phi_{L} = 0$).
Specifically, the curl--free nature of $\mathbf{F}_{L}$ follows from its definition as the gradient of a scalar, while
its divergence--free nature requires that $\phi_{L}$ satisfies the Laplace equation. As a harmonic field,
$\phi_{L}$ must obey the mean value theorem, i.e. that its value at any point ${\mathbf{x}}$ is equal to its average on any
spherical surface of arbitrary radius surrounding ${\mathbf{x}}$. This means that $\phi_{L}$ can contain no {\it local}
maxima or minima, and therefore that any maxima/minima must occur on its outer boundary. The properties of $\phi_{L}$
are therefore determined entirely by boundary conditions: $\mathbf{F}_{L} = {\nabla}\phi_{L}$ will quantify 
large-scale, smooth gradients in $\mathbf{F}$ that cannot be assigned to either ${\mathbf{F}}_{\perp}$ or ${\mathbf{F}}_{||}$.

For our study here, we use numerically simulated fields (with the momentum density ${\rho}{\mathbf{v}}$ playing the role of $\mathbf{F}$)
that obey periodic boundary conditions. For these fields, the multiplicity of choices for the ``boundary'' and the condition of
no local maxima/minima ensure that $\phi_{L}$ is a constant and therefore that $\mathbf{F}_{L} = 0$, and the Helmholtz Decompositon
is unique. 
For real momentum fields encountered in the interstellar medium (ISM), we must rely on finding clouds that are sufficiently isolated that boundary conditions
on ${\rho}{\mathbf{v}}$ are not a concern. Our method is therefore best-suited to molecular clouds that are bounded in space (i.e.
by a fall--off in density $\rho$ that ensures no non-zero ${\rho}{\mathbf{v}}$ values on their boundary).  

The more widely-distributed atomic medium is less suited to application of our method, since it will be difficult to ensure
the absence of large-scale gradients in any finite field. Finally, it should be noted that ${\rho}{\mathbf{v}}$ will be continuous
across the atomic/molecular transition and restriction to the molecular component is a necessarily limited description of the
ISM as a single fluid -- not to mention restriction to practically-observable regions using trace molecules, and the possible
influence of inter-mixed atomic/molecular zones. However, these problems are not in principle insurmountable (with sufficient data)
though they may pose considerable challenges if a complete description of the ISM fluid is desired.

\subsection{Projection from 3D to 2D}

Consider the case where we have access to only one scalar component of the vector field ${\mathbf{F}}$. 
We assume that the accessible component is oriented along the line-of-sight (as in, for example, a spectral line
observation), which we take as the $z$-direction.
The observable component of ${\mathbf{F}}$ is then $F_{z} = F_{z\perp} + F_{z||}$ where $F_{z\perp}$ and $F_{z||}$ are the $z$ components of the transverse
and longitudinal parts of ${\mathbf{F}}$. The contribution of $F_{z}$ to the Fourier transformed field -- $\tilde{\mathbf{F}}(\mathbf{k})$ --
is $\tilde{F}_{z}$ and this is oriented along the $k_{z}$-direction (i.e. $F_{z} \hat{\mathbf{z}}$ transforms 
into $\tilde{F}_{z} \hat{\mathbf{k}}_{z}$ where $\hat{\mathbf{z}}$ and $\hat{\mathbf{k}}_{z}$ are unit vectors in the $z$
and $k_{z}$ directions respectively).

\begin{figure*}
\includegraphics[width=84mm]{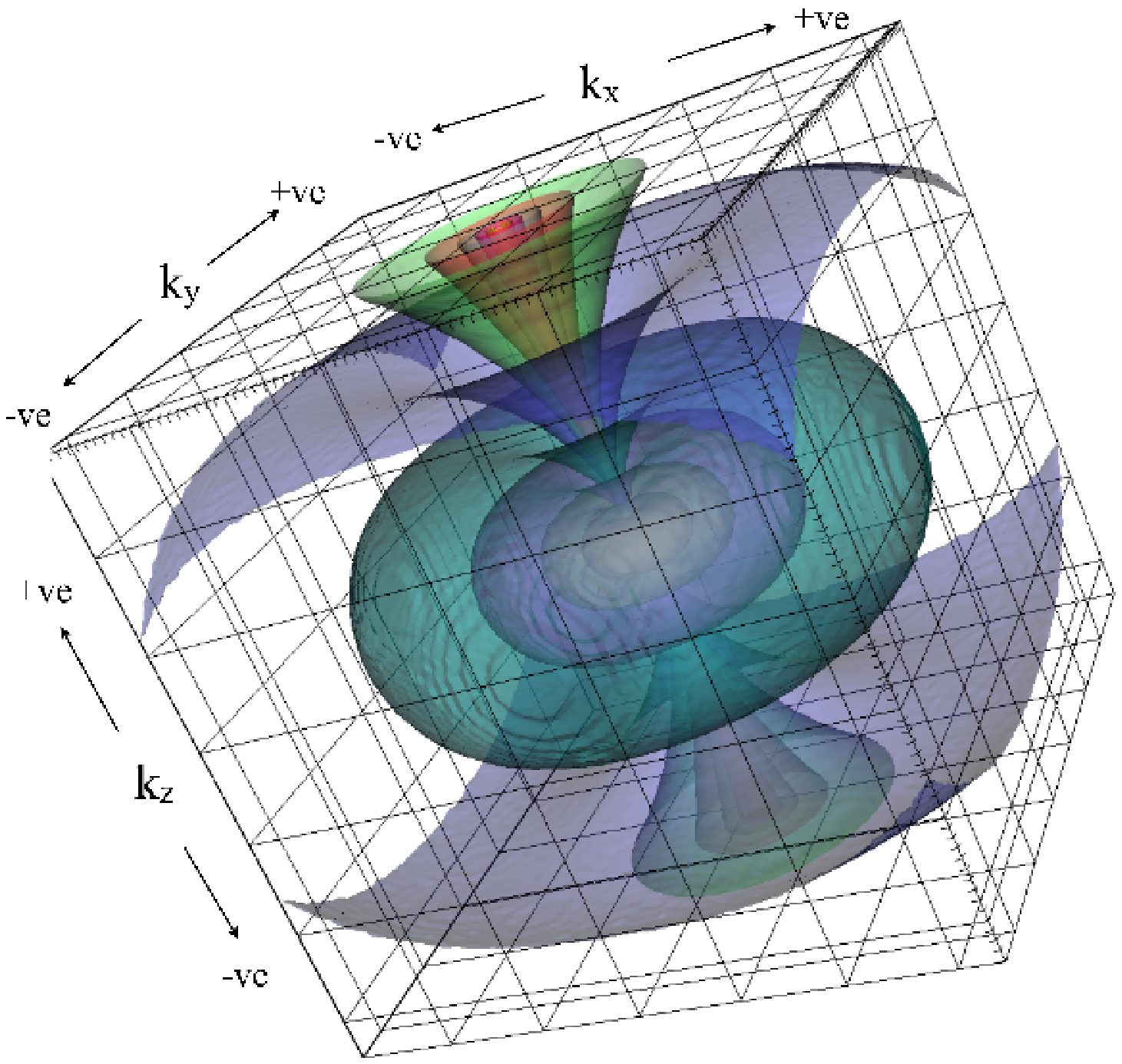}
\includegraphics[width=84mm]{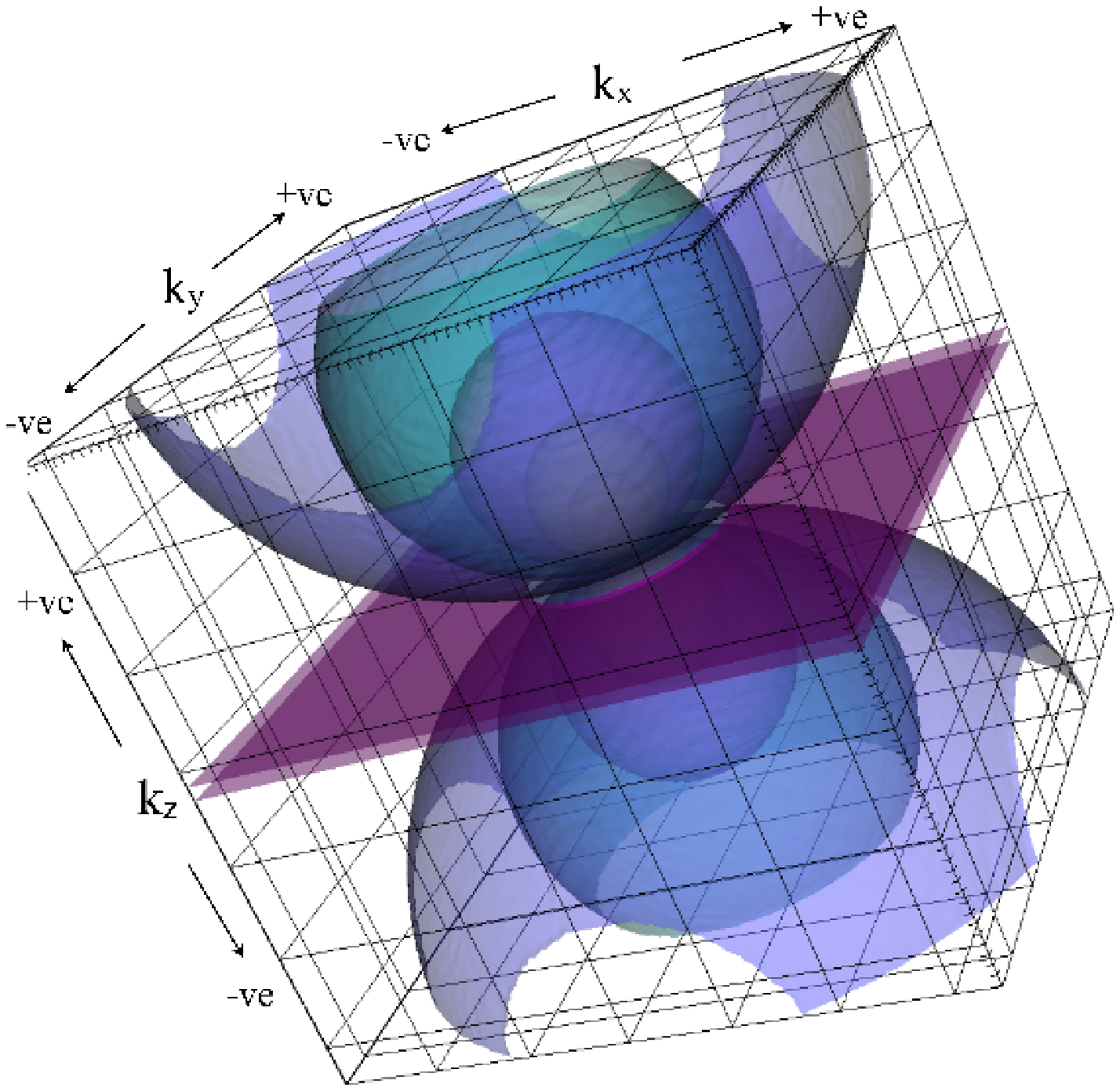}
\caption{3D renderings of constant power surfaces in the power spectra of $F_{z\perp}$ (left) and $F_{z_{||}}$ (right),
given by Equation~(\ref{fztin3d})
and Equation~(\ref{power1}) respectively. Isotropic power spectra: $\tilde{{\mathbf{F}}}_{\perp} \cdot
\tilde{{\mathbf{F}}}_{\perp}^{*} \propto k^{-4}$, and $\tilde{{\mathbf{F}}}_{||} \cdot \tilde{{\mathbf{F}}}_{||}^{*} \propto k^{-4}$ have been assumed. In both
panels, the origin of coordinates ($\mathbf{k} = 0$) lies in the
centre of the image.}
\label{fig:powplot0}
\end{figure*}

The condition $\mathbf{k} \cdot \tilde{\mathbf{F}}_{\perp} = 0$
requires that $\tilde{F}_{z\perp} = 0$ along the $k_{z}$ axis (where $k_{x} = k_{y} = 0$). Clearly  
we must also find that $\tilde{F}_{z\perp} = \tilde{F}_{z}$ everywhere in the plane $k_{z} = 0$, since there
the condition $\mathbf{k} \times \tilde{\mathbf{F}}_{||} = 0$ requires that $\tilde{F}_{z||} = 0$ in this plane.

In this paper, we only consider fields for which the transverse and longitudinal fields, ${\mathbf{F}}_{\perp}$ and
${\mathbf{F}}_{||}$, are statistically isotropic, i.e. that their power spectra ($\tilde{{\mathbf{F}}}_{\perp} \cdot \tilde{{\mathbf{F}}}_{\perp}^{*}$ and
$\tilde{{\mathbf{F}}}_{||} \cdot \tilde{{\mathbf{F}}}_{||}^{*}$) may be written as functions of $k=|\mathbf{k}|$ alone,
with no explicit angular dependence in Fourier space. In this case, we may write:
\begin{equation}
\tilde{{\mathbf{F}}}_{\perp} \cdot \tilde{{\mathbf{F}}}_{\perp}^{*} = F^{2}_{{\perp}0} f_{\perp}(k) ,
\label{ftkiso}
\end{equation}
\begin{equation}
\tilde{{\mathbf{F}}}_{||} \cdot \tilde{{\mathbf{F}}}_{||}^{*} = F^{2}_{||0} f_{||}(k) ,
\end{equation}
where $ F^{2}_{{\perp}0}$ and $ F^{2}_{||0}$ are scaling factors and $f_{\perp}(k)$ and $f_{||}(k)$ describe
the $k$-dependent power distributions. Note that for such isotropic fields the transverse
and longitudinal power distributions for a single scalar component are not isotropic, but have
predictable anisotropic structure determined by the following equations:
\begin{equation}
\tilde{F}_{z||} \tilde{F}_{z||}^{*} = \tilde{{\mathbf{F}}}_{||} \cdot \tilde{{\mathbf{F}}}_{||}^{*} \frac{k^{2}_{z}}{k^{2}} ,
\label{power1}
\end{equation}
\begin{equation}
\tilde{F}_{z\perp} \tilde{F}_{z\perp}^{*} =  \tilde{{\mathbf{F}}}_{\perp} \cdot \tilde{{\mathbf{F}}}_{\perp}^{*} \frac{k^{2}_{x}+k^{2}_{y}}{2k^{2}} ,
\label{fztin3d}
\end{equation}
\begin{equation}
\tilde{F}_{x||} \tilde{F}_{x||}^{*} = \tilde{{\mathbf{F}}}_{||} \cdot \tilde{{\mathbf{F}}}_{||}^{*} \frac{k^{2}_{x}}{k^{2}} ,
\end{equation}
\begin{equation}
\tilde{F}_{x\perp} \tilde{F}_{x\perp}^{*} = \tilde{{\mathbf{F}}}_{\perp} \cdot \tilde{{\mathbf{F}}}_{\perp}^{*} \frac{k^{2}_{y}+k^{2}_{z}}{2k^{2}} ,
\label{fxtin3d}
\end{equation}
\begin{equation}
\tilde{F}_{y||} \tilde{F}_{y||}^{*} = \tilde{{\mathbf{F}}}_{||} \cdot \tilde{{\mathbf{F}}}_{||}^{*} \frac{k^{2}_{y}}{k^{2}} ,
\end{equation}
\begin{equation}
\tilde{F}_{y\perp} \tilde{F}_{y\perp}^{*} = \tilde{{\mathbf{F}}}_{\perp} \cdot \tilde{{\mathbf{F}}}_{\perp}^{*} \frac{k^{2}_{x}+k^{2}_{z}}{2k^{2}} .
\label{power2}
\end{equation}
It is worth briefly mentioning here that, as can be checked in the above equations, 
the individual scalar components of the vector
field are only fully isotropic if, at each ${\mathbfit{k}}$, we have 
$\tilde{{\mathbf{F}}}_{\perp}$~$\cdot$~$\tilde{{\mathbf{F}}}_{\perp}^{*}$~$=$~$2\tilde{{\mathbf{F}}}_{||} \cdot \tilde{{\mathbf{F}}}_{||}^{*}$.
This is the case for (e.g.) fractional Brownian motion (fBm) fields - i.e. there is twice as
much power in the transverse component as the longitudinal component. In general, this will not
be true.

To illustrate the above power spectra, we show example 3D renderings of $\tilde{F}_{z\perp} \tilde{F}_{z\perp}^{*}$
and $\tilde{F}_{z_{||}} \tilde{F}_{z_{||}}^{*}$ in Figure~\ref{fig:powplot0}. Isotropic power spectra
of $\tilde{{\mathbf{F}}}_{\perp} \cdot \tilde{{\mathbf{F}}}_{\perp}^{*} \propto k^{-4}$, and 
$\tilde{{\mathbf{F}}}_{||} \cdot \tilde{{\mathbf{F}}}_{||}^{*} \propto k^{-4}$ have been assumed. 
The power spectrum of the longitudinal component of $F_{z}$ has a characteristic ``hourglass'' appearance,
resulting from the suppression (nulling) of power near (at) $k_{z} = 0$. Since ${\tilde{\mathbf{F}}}_{||}$ is aligned
with ${\mathbf{k}}$, this means that $\tilde{F}_{z_{||}}$ must be zero in the $k_{z} = 0$ plane. Conversely,
the transverse power is maximized (at fixed $k$) in the plane $k_{z} = 0$, and diminishes as $|k_{z}/k|$
approaches unity. An instructive reference point can be obtained by considering the power distributions
along the line ($k_{x} = 0, k_{y} = 0, k_{z} \neq 0$). Here, the longitudinal power is entirely
contained in the $z$-component, while the transverse power is equally split (assuming isotropy) between
the $x$-component and the $y$-component (Equations~(\ref{fxtin3d}) and (\ref{power2}) with $k_{x} = k_{y} = 0$).

The power spectra of the other ($x, y$) components share the same form, but with different
orientations, such that (as we assume here) the total transverse and longitudinal power spectra are
isotropic (functions of $k$ alone):
\begin{equation}
\tilde{{\mathbf{F}}}_{\perp} \cdot \tilde{{\mathbf{F}}}_{\perp}^{*} = \tilde{F}_{x\perp} \tilde{F}_{x\perp}^{*} + \tilde{F}_{y\perp} \tilde{F}_{y\perp}^{*} + \tilde{F}_{z\perp} \tilde{F}_{z\perp}^{*} = F^{2}_{{\perp}0}f_{\perp}(k) ,
\end{equation} 
\begin{equation}
\tilde{{\mathbf{F}}}_{||} \cdot \tilde{{\mathbf{F}}}_{||}^{*} = \tilde{F}_{x||} \tilde{F}_{x||}^{*} + \tilde{F}_{y||} \tilde{F}_{y||}^{*} + \tilde{F}_{z||} \tilde{F}_{z||}^{*} = F^{2}_{{||}0}f_{||}(k) ,
\end{equation}
as can be checked by Equations~(\ref{ftkiso}--\ref{power2}). In practice, power spectra, whether observed or simulated,
will not be {\it precisely} isotropic but we assume that
they are statistically isotropic--i.e. that the power values fluctuate randomly around the
form assumed above. Note that if the isotropy is only statistical, the geometry still requires
that $\tilde{F}_{z}$ has no longitudinal power in the $k_{z} = 0$ plane.

An important property of the above power distributions is that in the plane $k_{z} = 0$: 
\begin{equation}
\tilde{F}_{z} \tilde{F}_{z}^{*} = \tilde{F}_{z\perp} \tilde{F}_{z\perp}^{*} =\frac{1}{2} \tilde{{\mathbf{F}}}_{\perp} \cdot \tilde{{\mathbf{F}}}_{\perp}^{*} .
\end{equation}
It is straightforward to show that if $F_{z}$ is spatially averaged over the $z$-axis via:
\begin{equation}
F_{z,p}(x,y) = \frac{1}{L} \displaystyle \int_{-L/2}^{L/2} {\mathrm{d}}z \; F_{z}(x, y, z) ,
\label{zproject}
\end{equation}
then the
Fourier transform of the line-of-sight averaged field, $F_{z,p}$, is given by:
\begin{equation}
\tilde{F}_{z,p}(k_{x},k_{y}) = \frac{1}{L} \tilde{F}_{z}(k_{x},k_{y},k_{z}=0)
\end{equation}
(see BFP ). 
This means that the projected field, $F_{z,p}$, contains only contributions from transverse
structure in 3D, since its Fourier transform is directly proportional to a $k_{z} = 0$ cut
through the 3D Fourier transform of $F_{z}$ (which contains only transverse contributions).
Therefore, its power spectrum is a direct measure of the $k_{z} = 0$ plane of
the 3D power spectrum of $F_{z\perp}$ (and, indeed, of ${\mathbf{F}}_{\perp}$). The power spectrum,
as a representation of the variance of $F_{z,p}$, can be used to obtain an estimate of the
variance of $F_{z\perp}$ in 3D, if the power spectra conform to the structures 
given in Equations~(\ref{power1}--\ref{power2}).
Explicitly, the power spectrum of $F_{z,p}$ is, for $k_{z} = 0$:
\begin{eqnarray}
\tilde{F}_{z,p} \tilde{F}_{z,p}^{*} (k_{x}, k_{y}) \lefteqn{ = \frac{1}{L^{2}}\tilde{F}_{z\perp} \tilde{F}_{z\perp}^{*} (k_{x}, k_{y}, k_{z}=0)} \nonumber \\
\lefteqn{ = \frac{1}{L^{2}} \frac{\tilde{{\mathbf{F}}}_{\perp} \cdot \tilde{{\mathbf{F}}}_{\perp}^{*}}{2} (k_{x}, k_{y}, k_{z}=0)}.
\label{projfz}
\end{eqnarray}

Using Parseval's Theorem, and referring to Equation~(\ref{fztin3d}), the variance of $F_{z,p}$ is given by:
\begin{equation}
\sigma_{F_{z,p}}^{2}= {\frac{1}{L^{4}}} \displaystyle\sum_{k_{x}=-\infty}^{\infty} \sum_{k_{y}=-\infty}^{\infty} \; \tilde{F}_{z,p} \tilde{F}_{z,p}^{*}  ,
\end{equation}
or making use of Equation~(\ref{projfz}):
\begin{equation}
\sigma_{F_{z,p}}^{2}= {\frac{1}{L^{6}}} \displaystyle\sum_{k_{x}=-\infty}^{\infty} \sum_{k_{y}=-\infty}^{\infty} \; \frac{\tilde{{\mathbf{F}}}_{\perp} \cdot \tilde{{\mathbf{F}}}_{\perp}^{*}}{2} .
\end{equation}
The variance of $F_{z\perp}$ in three dimensions is given by:
\begin{equation}
\sigma_{F_{z\perp}}^{2} = {\frac{1}{L^{6}}} \displaystyle\sum_{k_{x}=-\infty}^{\infty} \sum_{k_{y}=-\infty}^{\infty} \sum_{k_{z}=-\infty}^{\infty} \; \tilde{{\mathbf{F}}}_{\perp} \cdot \tilde{{\mathbf{F}}}_{\perp}^{*} \frac{k^{2}_{x}+k^{2}_{y}}{2k^{2}} ,
\end{equation}
so that, under the assumption of isotropy (i.e. Equation~(\ref{ftkiso})), we may then use the measured
variance of $F_{z,p}$ to estimate the variance of $F_{z\perp}$ in three dimensions via:
\begin{equation}
\sigma_{F_{z\perp}}^{2} = \sigma_{F_{z,p}}^{2} \times \frac{\displaystyle\sum_{k_{x}=-\infty}^{\infty} \sum_{k_{y}=-\infty}^{\infty} \sum_{k_{z}=-\infty}^{\infty} \; f_{\perp}(k) \frac{k^{2}_{x}+k^{2}_{y}}{k^{2}}}{\displaystyle\sum_{k_{x}=-\infty}^{\infty} \sum_{k_{y}=-\infty}^{\infty} \; f_{\perp}(k)} ,
\label{mastereq}
\end{equation}
where we note that $f_{\perp}(k)$ can be directly measured using the power spectrum of the projected
field, $\tilde{F}_{z,p} \tilde{F}_{z,p}^{*}$.
The overall scaling of the fields (controlled by $F_{{\perp}0}$ in Equation~(\ref{ftkiso})) is unimportant
in determining the {\it ratio} $\sigma_{F_{z\perp}}^{2}/\sigma_{F_{z,p}}^{2}$ but {\it is} important if the absolute variance
$\sigma_{F_{z\perp}}^{2}$ is desired. Even more straightforwardly, noting that for an isotropic field,
$\sigma_{F_{z\perp}}^{2} = \sigma_{{\mathbf{F}}_{\perp}}^{2}/3$, we can also write:
\begin{equation}\sigma_{F_{z\perp}}^{2} = \frac{2}{3} \sigma_{F_{z,p}}^{2} \times \frac{\displaystyle\sum_{k_{x}=-\infty}^{\infty} \sum_{k_{y}=-\infty}^{\infty} \sum_{k_{z}=-\infty}^{\infty} \; f_{\perp}(k) }{\displaystyle\sum_{k_{x}=-\infty}^{\infty} \sum_{k_{y}=-\infty}^{\infty} \; f_{\perp}(k)} .
\label{mastereqmod}
\end{equation}

If, in addition, a measurement of the total variance of $F_{z}$ is available, then the fractional power in
transverse modes can be calculated. Even if only the ratio of projected-to-total variance, $\sigma_{F_{z,p}}^{2}/\sigma_{F_{z}}^{2}$, 
is known, the fractional power in transverse modes can still be calculated via:
\begin{equation}
\frac{\sigma_{F_{z\perp}}^{2}}{\sigma_{F_{z}}^{2}}  = \frac{\sigma_{F_{z,p}}^{2}}{{\sigma_{F_{z}}^{2}}}  \times \frac{\displaystyle\sum_{k_{x}=-\infty}^{\infty} \sum_{k_{y}=-\infty}^{\infty} \sum_{k_{z}=-\infty}^{\infty} \; f_{\perp}(k) \frac{k^{2}_{x}+k^{2}_{y}}{k^{2}}}{\displaystyle\sum_{k_{x}=-\infty}^{\infty} \sum_{k_{y}=-\infty}^{\infty} \; f_{\perp}(k)} ,
\label{mastereqA}
\end{equation}
or:
\begin{equation}
\frac{\sigma_{F_{z\perp}}^{2}}{\sigma_{F_{z}}^{2}}  = \frac{2}{3} \frac{\sigma_{F_{z,p}}^{2}}{{\sigma_{F_{z}}^{2}}}  \times \frac{\displaystyle\sum_{k_{x}=-\infty}^{\infty} \sum_{k_{y}=-\infty}^{\infty} \sum_{k_{z}=-\infty}^{\infty} \; f_{\perp}(k) }{\displaystyle\sum_{k_{x}=-\infty}^{\infty} \sum_{k_{y}=-\infty}^{\infty} \; f_{\perp}(k)} ,
\label{mastereqAmod}
\end{equation}
and again under the assumption of isotropy, this should be equal to the fractional power in transverse
modes for the full vector field ${\mathbf{F}}$, i.e.:
\begin{equation}
\frac{\sigma_{F_{\perp}}^{2}}{\sigma_{F}^{2}} \approx \frac{\sigma_{F_{z\perp}}^{2}}{\sigma_{F_{z}}^{2}} .
\label{1to3D}
\end{equation}
Note that there is no requirement that the longitudinal 
power spectrum to be known, nor be in any way dependent on the transverse power spectrum. 

A simpler version of Equation~(\ref{mastereq}) was used by BFP to estimate
the variance of normalised density, $\sigma^{2}_{\rho/\rho_{0}}$ from the observationally accessible normalised
column density variance $\sigma^{2}_{N/N_{0}}$ (where $\rho$ and $N$ are density and column density, and
$\rho_{0}$ and $N_{0}$ are their mean values, respectively). The appropriately modified form of Equation~(\ref{mastereq}) for 
this purpose is:
\begin{equation}
\sigma^{2}_{\rho/\rho_{0}} = \sigma^{2}_{N/N_{0}} \times \frac{\left(\displaystyle\sum_{k_{x}=-\infty}^{\infty} \sum_{k_{y}=-\infty}^{\infty} \sum_{k_{z}=-\infty}^{\infty} \; f(k)\right) - f(0)}{\left(\displaystyle\sum_{k_{x}=-\infty}^{\infty} \sum_{k_{y}=-\infty}^{\infty} \; f(k)\right) - f(0)} ,
\label{mastereq2}
\end{equation}
where it was assumed that the density power spectrum was isotropic, described by the function $f(k)$, and the mean value of the fields
(i.e. the zero-frequency component $f(0)$) is explicitly subtracted from the summations. Above we had assumed that a frame could be
chosen in which $\langle {\mathbf{F}} \rangle = 0$, whereas BFP used the positive-definite nature of $\rho$ and $N$ to provide a suitable normalisation. 
Obviously, Equation~(\ref{mastereq2}) does not include the $(k^{2}_{x}+k^{2}_{y})/k^{2}$
factor either, since $\rho$ is a scalar and not subject to the vector projection factors that determine $F_{z\perp}$.
For practical application of Equations~(\ref{mastereq}-\ref{mastereq2}) the sums only extend to the maximum
wavevector observable in the field, the consequences of which are discussed in BFP and Brunt (2010).

\subsection{Observational Considerations}

We now review the information provided by spectral line observations of the interstellar medium (ISM), with a view
to determining which physical field(s) may be analysed by the above system. The principal requirements for such
a field are (1) that it is a 3D vector field, and (2) that one of its components can be projected (or averaged) over 
the line-of-sight with no weighting by other variable physical fields. It turns out, as explained below, that the
momentum density field, ${\mathbfit{p}} = \rho {\mathbfit{v}}$, (hereafter simply ``momentum'') most closely satisfies these requirements. Application
of the method to velocity fields is impossible except under conditions of uniform density, which are essentially
never encountered in the ISM. Below we therefore develop a scheme whereby the fraction of momentum power
in transverse modes may be estimated observationally -- i.e. evaluation of Equation~(\ref{mastereqA}).

Here we only consider an optically thin isothermal medium with uniform excitation, so that the 
infinitesimal contribution to the spectral line intensity
generated by density $\rho$ along an infinitesimal path length ${\mathrm{d}}z$ at position $z$ 
is given by:
\begin{equation}
{\mathrm{d}}I(v) = e \rho \phi(v - v_{z}(z)) {\mathrm{d}}z , 
\end{equation}
where $e$ is a constant
and $\phi(v - v_{z})$ is the normalised profile function, which can usually be represented by a Gaussian:
\begin{equation}
\phi(v - v_{z}) = \frac{1}{\sqrt{2 {\mathrm{\pi}} \sigma^{2}_{t,i}}} \exp \left( - \frac{(v - v_{z})^{2}}{2\sigma^{2}_{t,i}} \right),
\end{equation}
where $\sigma^{2}_{t,i}$ represents the dispersion caused by thermal and instrumental broadening.
Integrating along the line-of-sight, the observed spectral line intensity is then:
\begin{equation}
I(x, y, v) = e \displaystyle \int_{-L/2}^{L/2} {\mathrm{d}}z \; \rho(x, y, z) \phi(v - v_{z}(x, y, z)) ,
\label{dwvhist}
\end{equation}
where we have assumed the emitting medium lies within a region of spatial size $L$.
For molecular spectral line observations,
$\sigma^{2}_{t,i}$ is usually very small compared to the overall velocity dispersion, so that a reasonable approximation
is $\phi(v - v_{z}) = \delta(v - v_{z})$ where $\delta(v - v_{z})$ is the Dirac delta function. In this
case, Equation~(\ref{dwvhist}) effectively describes the intensity as a ``density-weighted histogram'' of line-of-sight
velocity, and indeed this is a standard method to represent optically-thin spectral line observations of
numerical simulations (e.g. Falgarone et al 1994; Ostriker, Stone, \& Gammie 2001). (Note that the delta function may be satisfied multiple times along the line-of-sight
for a turbulent medium.)

With $\phi(v - v_{z}) = \delta(v - v_{z})$, consider the observationally-accessible integral (the
first velocity-moment of the intensity):
\begin{eqnarray}
W_{1}(x, y)\lefteqn{= \displaystyle \int_{-\infty}^{\infty} {\mathrm{d}}v \; I(x, y, v) v} \nonumber \\
\lefteqn{= \displaystyle \int_{-\infty}^{\infty} {\mathrm{d}}v \; e \displaystyle \int_{-L/2}^{L/2} {\mathrm{d}}z \; \rho(x, y, z) \delta(v - v_{z}) v} \nonumber \\
\lefteqn{= e \displaystyle \int_{-L/2}^{L/2} {\mathrm{d}}z \; \rho(x, y, z)  v_{z}(x, y, z)} \nonumber \\
\lefteqn{= e \displaystyle \int_{-L/2}^{L/2} {\mathrm{d}}z \;  p_{z}(x, y, z)} \nonumber \\
\lefteqn{= e L p_{z,p} ,}
\label{pzpeq}
\end{eqnarray}
where $p_{z} = \rho v_{z}$ is the $z$-component of the momentum, and $p_{z,p}$ is (c.f. Equation~(\ref{zproject})):
\begin{equation}
p_{z,p}(x, y) = \frac{1}{L} \int_{-L/2}^{L/2} {\mathrm{d}}z \; p_{z}(x, y, z) .
\end{equation}
The momentum field therefore satisfies the spatial projection requirement (up to constants of proportionality), while
the velocity field does so only under conditions of uniform density. Therefore in equation~(\ref{mastereqA})
we will set $F_{z} = p_{z}$ and $F_{z,p} = p_{z,p}$.

At this point, we identify the following ratio: $\sigma^{2}_{p_{z,p}}/\sigma^{2}_{p_{z}}$ -- i.e. the 
fraction of $z$-momentum power (variance) projected into 2D -- as the most relevant quantity to estimate 
observationally, since measurements of $\sigma^{2}_{p_{z,p}}/\sigma^{2}_{p_{z}}$ and $f_{\perp}(k)$ are needed 
for the evaluation of equation~(\ref{mastereqA}). We already have $f_{\perp}(k)$, as this can be derived from the
angular average of the power spectrum of $W_{1}(x, y)$ (note that the overall normalisation is unimportant).

While $e$ and $L$ (in equation~(\ref{pzpeq})) can in principle be estimated, a better procedure is to normalise them out. The
integrated intensity (the zeroth velocity-moment of the intensity) is:
\begin{equation}
W_{0}(x,y ) = \displaystyle \int_{-\infty}^{\infty} {\mathrm{d}}v \; I(x, y, v)  = e L \rho_{p}(x, y) = eN,
\end{equation}
where 
\begin{equation}
\rho_{p}(x, y) = \frac{1}{L} \int_{-L/2}^{L/2} {\mathrm{d}}z \; \rho(x, y, z) 
\end{equation}
is the line-of-sight average of $\rho$ ($N = L\rho_{p}$ is the column density).

By computing spatial averages over $x$ and $y$ (denoted by angle brackets) the ratio:
\begin{equation}
\frac{\langle W^{2}_{1} \rangle}{\langle W^{2}_{0} \rangle} = \frac{\sigma^{2}_{p_{z,p}}}{\langle \rho^{2}_{p} \rangle}
\end{equation}
can be formed, where we have assumed the calculations are done in the zero-momentum frame ($\langle p_{z,p} \rangle = 0$,
or equivalently, $\langle W_{1} \rangle = 0$).

We review now what remains to be calculated in order to form the ratio $\sigma^{2}_{p_{z,p}}/\sigma^{2}_{p_{z}}$ needed
for application of Equation~(\ref{mastereqA}).
As is evident from the dimensionality of $\sigma^{2}_{p_{z,p}}/\langle \rho^{2}_{p} \rangle$, it 
turns out that a measurement of its 3D analogue, $\sigma^{2}_{p_{z}}/\langle \rho^{2} \rangle$, is the most useful way
to proceed. If this quantity was available, then we could form the ratio:
\begin{equation}
\frac{\sigma^{2}_{p_{z,p}}/\langle \rho_{p}^{2} \rangle}{\sigma^{2}_{p_{z}}/\langle \rho^{2} \rangle} = \frac{\sigma^{2}_{p_{z,p}}}{\sigma^{2}_{p_{z}}} \frac{\langle \rho^{2} \rangle}{\langle \rho_{p}^{2} \rangle} = \frac{\sigma^{2}_{p_{z,p}}}{\sigma^{2}_{p_{z}}} \frac{\langle (\rho/\rho_{0})^{2} \rangle}{\langle (N/N_{0})^{2} \rangle} ,
\label{halfway}
\end{equation}
where in the last step we have written $\rho = \rho_{0}(\rho/\rho_{0})$ and $\rho_{p} = N/L = \rho_{0} (N/\rho_{0} L) = \rho_{0} (N/N_{0})$ 
since the mean column density is $N_{0} = \rho_{0}L$.

The quantity $\langle (N/N_{0})^{2} \rangle$ can be directly calculated via:
\begin{equation}
\langle (N/N_{0})^{2} \rangle = \frac{\langle N^{2} \rangle}{\langle N_{0} \rangle^{2}} = \frac{\langle W^{2}_{0} \rangle}{\langle W_{0} \rangle^{2}} ,
\end{equation}
and we have a means to calculate $\langle (\rho/\rho_{0})^{2} \rangle$ using Equation~(\ref{mastereq2})
since $\langle (\rho/\rho_{0})^{2} \rangle =  1 + \sigma^{2}_{\rho/\rho_{0}}$  and
$\sigma^{2}_{N/N_{0}}  = \langle (N/N_{0})^{2} \rangle - 1$.
Note that $f(k)$ is the angular average of the column density power spectrum and since 
the overall normalisation is unimportant, the integrated intensity power spectrum, $\tilde{W}_{0} \tilde{W}^{*}_{0}$, can be used to measure this.

With $\langle (N/N_{0})^{2} \rangle$ and $\langle (\rho/\rho_{0})^{2} \rangle$ known, the 
desired ratio $\sigma^{2}_{p_{z,p}}/\sigma^{2}_{p_{z}}$ is given by (see Equation~(\ref{halfway})):
\begin{equation}
\frac{\sigma^{2}_{p_{z,p}}}{\sigma^{2}_{p_{z}}} = \left[\frac{\sigma^{2}_{p_{z,p}}}{\langle \rho^{2}_{p} \rangle} \right] \left[\frac{\langle (N/N_{0})^{2} \rangle}{\langle (\rho/\rho_{0})^{2} \rangle} \right] \left[ \frac{\sigma^{2}_{p_{z}}}{\langle \rho^{2} \rangle} \right]^{-1} .
\label{threequarterway}
\end{equation}

It still remains to find an observational estimate of 
$\sigma^{2}_{p_{z}}/\langle \rho^{2} \rangle$. This is a three-dimensional quantity -- the
ratio of mean squared $z$-momentum to mean squared density. Explicitly:
\begin{equation}
\frac{\sigma^{2}_{p_{z}}}{\langle \rho^{2} \rangle} = \frac{\frac{1}{L^{3}} \displaystyle \int_{-L/2}^{L/2} {\mathrm{d}}x \; \displaystyle \int_{-L/2}^{L/2} {\mathrm{d}}y \; \displaystyle \int_{-L/2}^{L/2} {\mathrm{d}}z \;  p_{z}^{2}}{\frac{1}{L^{3}} \displaystyle \int_{-L/2}^{L/2} {\mathrm{d}}x \; \displaystyle \int_{-L/2}^{L/2} {\mathrm{d}}y \; \displaystyle \int_{-L/2}^{L/2} {\mathrm{d}}z \; \rho^{2}} , 
\end{equation}
where we have again assumed the calculations are performed in the zero momentum frame.
From a slightly different perspective, we see that it may also be viewed as the $z$-velocity dispersion calculated
with a $\rho^{2}$ weight:
\begin{equation}
\frac{\sigma^{2}_{p_{z}}}{\langle \rho^{2} \rangle} = \frac{\frac{1}{L^{3}} \displaystyle \int_{-L/2}^{L/2} {\mathrm{d}}x \; \displaystyle \int_{-L/2}^{L/2} {\mathrm{d}}y \; \displaystyle \int_{-L/2}^{L/2} {\mathrm{d}}z \;  \rho^{2} v_{z}^{2}}{\frac{1}{L^{3}} \displaystyle \int_{-L/2}^{L/2} {\mathrm{d}}x \; \displaystyle \int_{-L/2}^{L/2} {\mathrm{d}}y \; \displaystyle \int_{-L/2}^{L/2} {\mathrm{d}}z \; \rho^{2}} = \frac{\langle \rho^{2} v^{2}_{z} \rangle}{\langle \rho^{2} \rangle}. 
\end{equation}
However, we only have access instead to the $z$-velocity dispersion calculated with a $\rho$ weight, as follows. 
Making use of the second velocity-moment of intensity, $W_{2}$:
\begin{eqnarray}
W_{2}(x, y)\lefteqn{= \displaystyle \int_{-\infty}^{\infty} {\mathrm{d}}v \; I(x, y, v) v^{2}} \nonumber \\
\lefteqn{= \displaystyle \int_{-\infty}^{\infty} {\mathrm{d}}v \; e \displaystyle \int_{-L/2}^{L/2} {\mathrm{d}}z \; \rho(x, y, z) \delta(v - v_{z}) v^{2}} \nonumber \\
\lefteqn{= e \displaystyle \int_{-L/2}^{L/2} {\mathrm{d}}z \; \rho(x, y, z)  v^{2}_{z}(x, y, z ) ,} 
\end{eqnarray}
a spatial average of $W_{2}$, normalised by the spatial average of $W_{0}$, gives:
\begin{eqnarray}
\frac{\langle W_{2} \rangle}{\langle W_{0} \rangle} \lefteqn{=  \frac{ \frac{e}{L^{2}} \displaystyle \int_{-L/2}^{L/2} {\mathrm{d}}x \; \displaystyle \int_{-L/2}^{L/2} {\mathrm{d}}y \; \displaystyle \int_{-L/2}^{L/2} {\mathrm{d}}z \; \rho(x, y, z)  v^{2}_{z}(x, y, z )}{ \frac{e}{L^{2}} \displaystyle \int_{-L/2}^{L/2} {\mathrm{d}}x \; \displaystyle \int_{-L/2}^{L/2} {\mathrm{d}}y \; \displaystyle \int_{-L/2}^{L/2} {\mathrm{d}}z \; \rho(x, y, z) }} \nonumber \\
\lefteqn{= \frac{\langle \rho v^{2}_{z} \rangle}{\langle \rho \rangle} .}
\label{w2overw0}
\end{eqnarray}
Thus we have a measure of $\langle \rho v^{2}_{z} \rangle/\langle \rho \rangle$ but require a measure of $\langle \rho^{2} v^{2}_{z} \rangle/\langle \rho^{2} \rangle$.
While there are circumstances in which $\langle \rho^{2} v^{2}_{z} \rangle/\langle \rho^{2} \rangle = \langle \rho v^{2}_{z} \rangle/\langle \rho \rangle$,
in general they will be different. They can be considered practically equal in two special cases: (1) a uniform or only very weakly varying density field, 
(2) statistically uncorrelated density and velocity fields. Neither of these conditions is expected to hold in the ISM, and corrective measures
are necessary. 
For now, we write, in advance of notation explained in the Appendix:
\begin{equation}
\frac{\sigma^{2}_{p_{z}}}{\langle \rho^{2} \rangle} = \frac{\langle \rho^{2} v^{2}_{z} \rangle}{\langle \rho^{2} \rangle} = g_{21} \frac{\langle \rho v^{2}_{z} \rangle}{\langle \rho \rangle} ,
\label{usingg21}
\end{equation}
where $g_{21}$ is a statistical correction factor of order unity. In the Appendix, we show
that $g_{21}$ may be written:
\begin{equation}
g_{21} = \langle \xi^{2} \rangle^{-\epsilon} ,
\end{equation}
where $\xi = \rho/\rho_{0}$ and $\epsilon$ is a small, positive constant. (We find for the numerical simulations
that $\epsilon$ is Mach number dependent.) In the Appendix, we discuss both numerical and observational
estimates of $\epsilon$ and look at the effects of $g_{21}$ on the analysis in Section~4.

Above, we have computed $\langle \rho v^{2}_{z} \rangle/\langle \rho \rangle$ via the observable ratio $W_{2}/W_{0}$, but
note that it may also be computed simply as the dispersion of the summed spectral line profile of the data. It is 
also important to recognise that a finite-width thermal/instrumental broadening term will cause overestimation of 
$\langle \rho v^{2}_{z} \rangle/\langle \rho \rangle$. The simple fix for this is to subtract
the thermal/instrumental dispersion from the raw measurement of $\langle \rho v^{2}_{z} \rangle/\langle \rho \rangle$
(i.e. the influence of finite thermal/instrumental broadening is simply to convolve the $\delta$-function-mapped
data with a smoothing kernel on the $v_{z}$-axis which will be typically Gaussian in form). We will return to this point, and other
practical observational considerations in Section~4 below.

Putting together all the above, equation~(\ref{threequarterway}) becomes:
\begin{equation}
\frac{\sigma^{2}_{p_{z,p}}}{\sigma^{2}_{p_{z}}} = \left[\frac{\sigma^{2}_{p_{z,p}}}{\langle \rho^{2}_{p} \rangle} \right] \left[\frac{\langle (N/N_{0})^{2} \rangle}{\langle (\rho/\rho_{0})^{2} \rangle} \right] \left[ g_{21} \frac{\langle \rho v^{2}_{z} \rangle}{\langle \rho \rangle} \right]^{-1} ,
\label{allofit}
\end{equation}
so that the fraction of $z$-momentum power in transverse modes is given by Equation~(\ref{mastereqA}):
\begin{equation}
\frac{\sigma^{2}_{p_{z\perp}}}{\sigma^{2}_{p_{z}}} =  \frac{\sigma_{p_{z,p}}^{2}}{{\sigma_{p_{z}}^{2}}}  \times \frac{\displaystyle\sum_{k_{x}=-k_{max}}^{k_{max}} \sum_{k_{y}=-k_{max}}^{k_{max}} \sum_{k_{z}=-k_{max}}^{k_{max}} \; f_{\perp}(k) \frac{k^{2}_{x}+k^{2}_{y}}{k^{2}}}{\displaystyle\sum_{k_{x}=-k_{max}}^{k_{max}} \sum_{k_{y}=-k_{max}}^{k_{max}} \; f_{\perp}(k)} ,
\label{allofit2}
\end{equation}
where $k_{max}$ is the maximum observed 1D wavenumber, and $f_{\perp}(k)$
is the angular average of the projected momentum power spectrum. Assuming isotropy,
equation~(\ref{allofit2}) provides an estimate of the ratio of transverse to total
momentum power in 3D, i.e.:
\begin{equation}
\frac{\sigma^{2}_{p_{\perp}}}{\sigma^{2}_{p}} \approx \frac{\sigma^{2}_{p_{z\perp}}}{\sigma^{2}_{p_{z}}} .
\label{allofitin3D}
\end{equation}

\subsection{Summary of observational measurements required}

Equations~(\ref{allofit}-\ref{allofitin3D}) can be written in terms of known observables:
\begin{equation}
\frac{\sigma^{2}_{p_{\perp}}}{\sigma^{2}_{p}} \approx \left[\frac{\langle W^{2}_{1} \rangle}{\langle W^{2}_{0} \rangle}\right] \left[ \frac{\langle W^{2}_{0} \rangle / \langle W_{0} \rangle^{2}}{1 + A(\langle W^{2}_{0} \rangle / \langle W_{0} \rangle^{2} - 1)} \right] \left[ g_{21} \frac{\langle W_{2} \rangle}{\langle W_{0} \rangle} \right]^{-1} B ,
\label{obseq}
\end{equation}
where:
\begin{equation}
A = \frac{\left(\displaystyle\sum_{k_{x}=-k_{max}}^{k_{max}} \sum_{k_{y}=-k_{max}}^{k_{max}} \sum_{k_{z}=-k_{max}}^{k_{max}} \; f(k)\right) - f(0)}{\left(\displaystyle\sum_{k_{x}=-k_{max}}^{k_{max}} \sum_{k_{y}=-k_{max}}^{k_{max}} \; f(k)\right) - f(0)} ,
\end{equation}
\begin{equation}
B = \frac{\displaystyle\sum_{k_{x}=-k_{max}}^{k_{max}} \sum_{k_{y}=-k_{max}}^{k_{max}} \sum_{k_{z}=-k_{max}}^{k_{max}} \; f_{\perp} (k) \frac{k^{2}_{x}+k^{2}_{y}}{k^{2}}}{\displaystyle\sum_{k_{x}=-k_{max}}^{k_{max}} \sum_{k_{y}=-k_{max}}^{k_{max}} \; f_{\perp}(k)} ,
\end{equation}
and:
\begin{equation}
f(k) = \frac{1}{2 {\mathrm{\pi}} k} \displaystyle \int_{0}^{2 {\mathrm{\pi}}} {\mathrm{d}}\phi \; \tilde{W}_{0}(k,\phi) \tilde{W}^{*}_{0}(k,\phi) ,
\end{equation}
\begin{equation}
f_{\perp}(k) = \frac{1}{2 {\mathrm{\pi}} k} \displaystyle \int_{0}^{2 {\mathrm{\pi}}} {\mathrm{d}}\phi \; \tilde{W}_{1}(k,\phi) \tilde{W}^{*}_{1}(k,\phi) ,
\end{equation}
are, respectively, the angular averages of the power spectra of the zeroth and first velocity moments 
of the spectral line intensities (in practice, computed by sums rather than integrals).

Finally, we note that in equation~(\ref{obseq}), it is evident that some factors cancel, suggesting further
simplification is possible. We have opted to leave it in the form presented since we believe it makes more
logical sense this way, and the individual terms (in square brackets) in the equation will be 
analysed below, along with the deprojection factor, $B$, in Section~4.

\section{Numerical Simulations}

\begin{figure*}
\includegraphics[width=84mm]{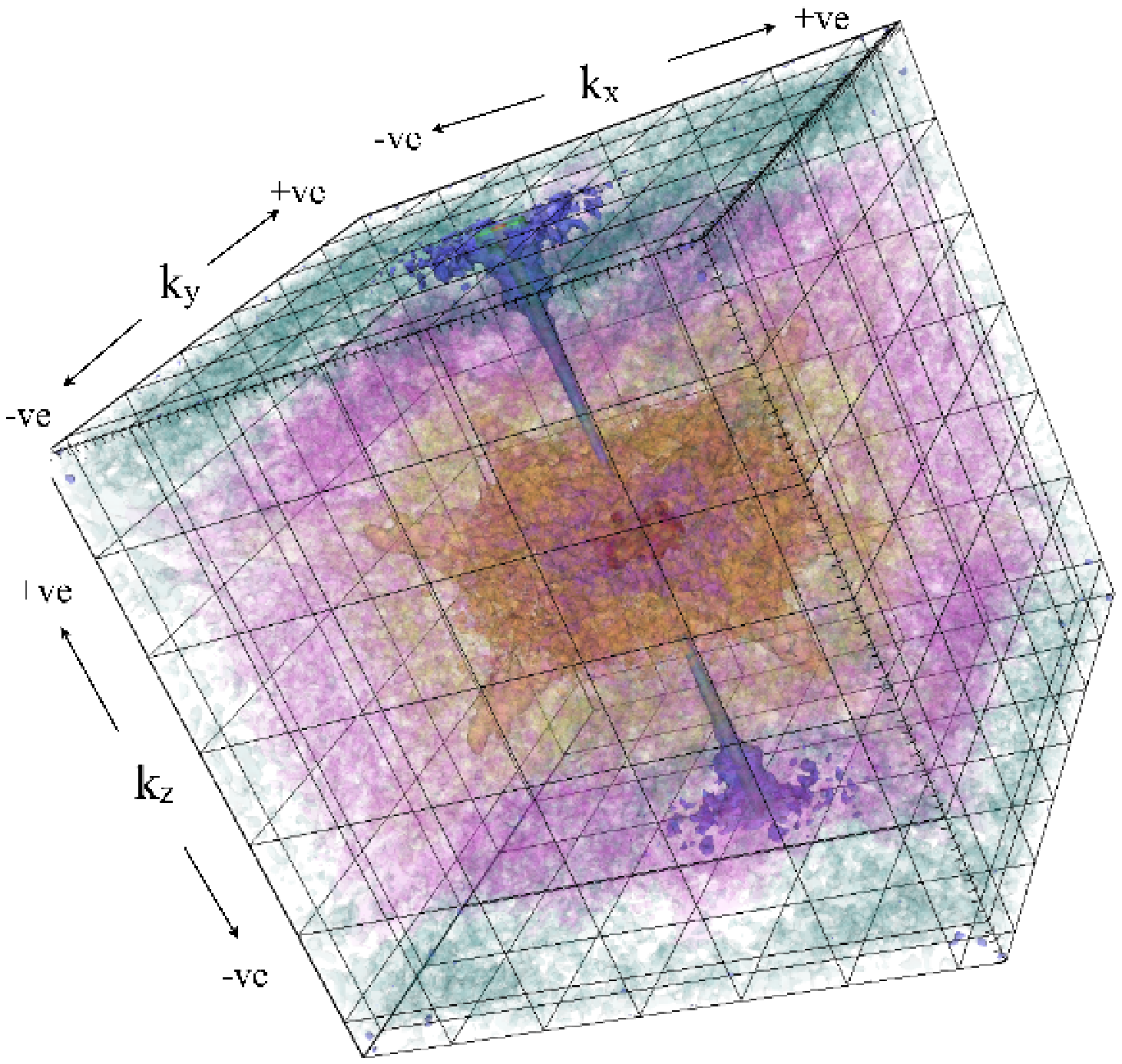}
\includegraphics[width=84mm]{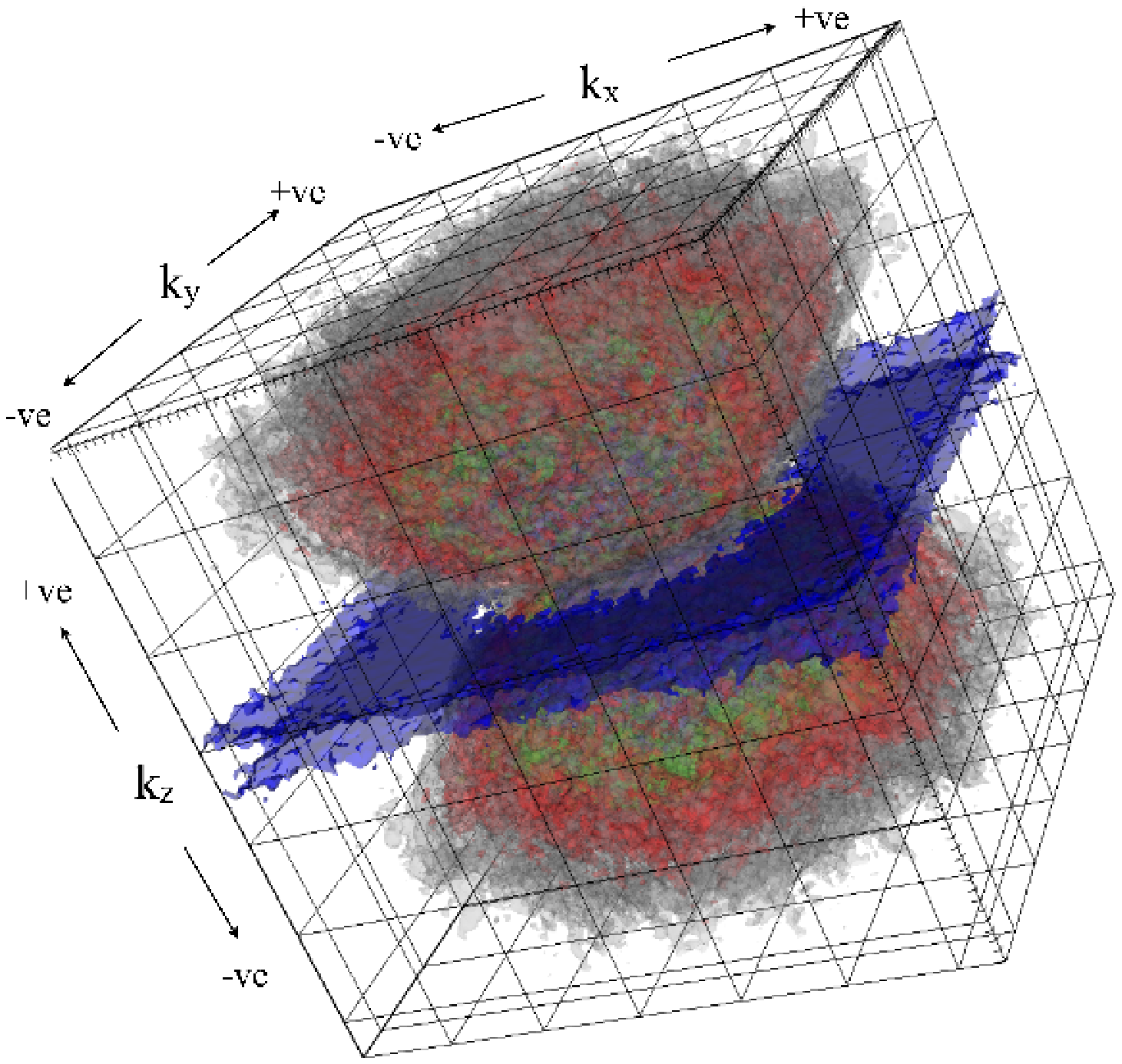}
\caption{3D renderings of {\bf constant power surfaces in} the power spectra of $p_{z\perp}$ (left) and $p_{z_{||}}$ (right) in a snapshot
taken from the numerical simulations (solenoidally-forced, rms Mach number = 5).
In both panels, the origin of coordinates ($\mathbf{k} = 0$) lies in the centre of the image.
(c.f. Figure~\ref{fig:powplot0}.)}
\label{fig:powplot00}
\end{figure*}

The simulations used to test the analytic method were performed with the astrophysical code
FLASH (Fryxell et al 2000; Dubey et el 2008), which integrates the ideal, three-dimensional,
magnetohydrodynamic (MHD) equations of compressible gas. The MHD equations were closed with a
polytropic equation of state, $P_\mathrm{th}=\cs^2\rho$, such that the gas remains isothermal
with a constant sound speed, $c_{s}$. We solve the MHD equations in the hydrodynamic limit (${\mathbf{B}} = 0$)
on three-dimensional, uniform grids with a fixed resolution of $256^3$ grid points and periodic boundary
conditions, using a positive-definite Riemann solver for ideal MHD
(Waagan, Federrath, \& Klingenberg 2011).

To drive turbulence, a stochastic forcing term ${\bf F_\mathrm{stir}}$ is applied as a source term in the
MHD momentum equation. Following common practice, the forcing only acts on large scales $1<k<3$
(where most of the power is injected at the $k=2$ mode in Fourier space, which corresponds to half
of the box size $L$ in physical space), i.e., the outer scale of molecular clouds, as favored by
observations (Ossenkopf \& Mac Low 2002; Brunt, Heyer, \& Mac Low 2009), such that turbulence develops
self-consistently on smaller scales. We use the Ornstein-Uhlenbeck (OU) process to model
${\bf F_\mathrm{stir}}$, which is a well-defined stochastic process with a finite autocorrelation
timescale (Eswaran \& Pope 1988;  Schmidt, Hillebrandt, \& Niemeyer 2006), leading to a smoothly varying
stochastic force field in space and time. Details about the OU process and the forcing applied in
this study can be found in Schmidt et al (2009), Federrath et al (2010),
and Konstandin et al (2012a). However, the essential point of our forcing approach is that we can
adjust the mixture of solenoidal and compressive modes of ${\bf F_\mathrm{stir}}$ arbitrarily.
This is achieved with the projection tensor $\mathcal{\underline{P}}^{\,\zeta}(\vect{k})$ in Fourier space.
In index notation, it reads
\begin{equation} \label{eq:projectionoperator}
\mathcal{P}_{ij}^\zeta = \zeta\,\mathcal{P}_{ij}^\perp+(1-\zeta)\,\mathcal{P}_{ij}^\parallel = \zeta\,\delta_{ij}+(1-2\zeta)\,\frac{k_i k_j}{|k|^2}\;,
\end{equation}
where $\delta_{ij}$ is the Kronecker symbol, and $\mathcal{P}_{ij}^\perp = \delta_{ij} - k_i k_j / k^2$ and
$\mathcal{P}_{ij}^\parallel = k_i k_j / k^2$ are the fully solenoidal and the fully compressive projection
operators, respectively. The ratio of compressive power to total power in ${\bf F_\mathrm{stir}}$ can be
derived from Equation~(\ref{eq:projectionoperator}) by evaluating the norm of the compressive component of
the projection tensor and dividing it by the total injected power, resulting in
\begin{equation} \label{eq:forcing_ratio}
\frac{F_\mathrm{comp}}{F_\mathrm{tot}} = \frac{(1-\zeta)^2}{1-2\zeta+3\zeta^2}\,,
\end{equation}
for three-dimensional space (Schmidt et al 2009; Federrath et al 2010). The projection
operator serves to construct a purely solenoidal force field by setting $\zeta=1$, while for $\zeta=0$, a
purely compressive force field is obtained. Any combination of solenoidal and compressive modes can be constructed
by choosing $\zeta\in[0,1]$. Here we compare simulations with $\zeta=1$ (sol) and $\zeta=0$ (comp).

Starting from a uniform density distribution and zero velocities, the forcing excites turbulent motions, which
approach a statistically steady state after about two turbulent crossing times,
$2T=L/(\mach\cs)$ (e.g., Klessen, Heitsch, \& Mac Low 2000; Klessen 2001; Heitsch, Mac Low, \& Klessen 2001;
Federrath, Klessen, \& Schmidt 2009; Federrath et al 2010; Price \& Federrath 2010; Micic et al 2012; Federrath 2013),
where $\mach$ denotes the three-dimensional, root-mean-squared sonic Mach number. We study simulations in both the
subsonic and supersonic regimes of turbulence with $\mach\sim0.1$, 0.5, 2, 5, and 15, each with the two limiting cases
of purely solenoidal and purely compressive forcing to basically cover the whole range of possible solenoidal momentum
ratios between zero and unity, in order to test the analytic method. To ensure that the initial transient phase
($t\lesssim 2T$) is not included in the following analysis, we consider snapshots at $t=3$, 4, and $5\,T$. Each of the
snapshots used in the analysis is thus separated by at least one crossing time, effectively representing statistically-independent
turbulent fields at this temporal separation. We thus improve the independent statistical sampling of our results by including
these three snapshots for each simulation, providing an estimate of the typical temporal variations of our results.

We also make use of previously conducted simulations for auxiliary information and testing purposes.
These simulations have been previously described in BFP and comprise smoothed particle hydrodynamics (SPH)
simulations of solenoidally forced turbulence at a range of (supersonic) Mach numbers and
fixed grid simulations of solenoidally-forced magnetohydrodynamic (MHD) turbulence at a range of
(supersonic) Mach numbers and a range of magnetic field strengths (see BFP for more details).

Finally in this Section, we compute example power spectra of longitudinal and transverse momentum components 
for comparison with Figure~\ref{fig:powplot0}. The Fourier transform ($\tilde{{\mathbf{p}}}_{||}$) of the 
longitudinal component of the momentum (${\mathbf{p}}_{||}$) is given by:
\begin{equation}
{\tilde{\mathbf{p}}}_{||} = {\mathbf{\hat{k}}}({\mathbf{\hat{k}}} \cdot \tilde{{\mathbf{p}}}) ,
\end{equation} 
where ${\mathbf{p}} = \rho {\mathbf{v}}$, and $\tilde{{\mathbf{p}}}$ is its Fourier transform.
The longitudinal momentum is then found by 
${\mathbf{p}}_{||} = $FT$^{-1}(\tilde{{\mathbf{p}}}_{||})$. We then extract the 
transverse component via:
\begin{equation}
{\mathbf{p}}_{\perp} = {\mathbf{p}} - {\mathbf{p}}_{||}.
\end{equation}
From the vector momenta, we can extract the $z$-components ($p_{z||}$ and $p_{z\perp}$) and measure
their power spectra. Figure~\ref{fig:powplot00} shows example 3D renderings for a solenoidally-driven
$p_{z}$-field at rms Mach number 5. The power spectra exhibit the same underlying symmetries as the power spectra
shown in Figure~\ref{fig:powplot0}, including the suppression (nulling) of longitudinal power as
$k_{z}$ approaches (equals) zero. (This geometrical aspect is enforced even if the power spectra
are anisotropic.)

\section{Application to Hydrodynamic Simulations}

\begin{figure}
\includegraphics[width=84mm]{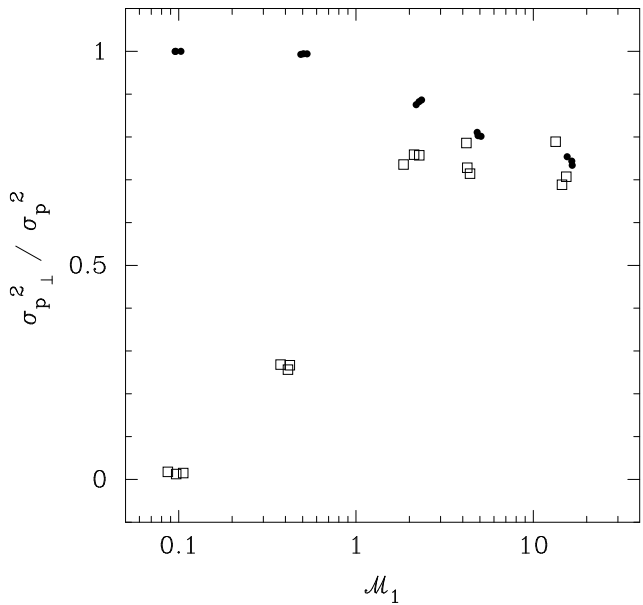}
\caption{Plot of the fraction of momentum power in transverse modes versus 3D density-weighted Mach number, ${\mathcal{M}}_{1}$. 
Dots: solenoidal forcing; open squares: compressive forcing.}
\label{fig:tfrac}
\end{figure}

We now apply the above method to the numerical simulations described in Section 3. Before doing so, we
examine the fraction of momentum power contained in transverse modes by direct calculation in 3D. These
fractions are calculated via Fourier space, making use of the conditions given in equations~(\ref{dot2})~and~(\ref{cross2}).
In Figure~\ref{fig:tfrac} we plot the ratio of transverse to total momentum power, $\sigma^{2}_{p_{\perp}} / \sigma^{2}_{p}$,
against the density-weighted Mach number, ${\mathcal{M}_{1}}$. At low Mach number (${\mathcal{M}_{1}} << 1$), 
the different forcing methods result in very different values of $\sigma^{2}_{p_{\perp}} / \sigma^{2}_{p}$ -- essentially 
all the power remains in the respective forcing modes. As ${\mathcal{M}_{1}}$ increases, the transverse
fractions appear to converge to $\sigma^{2}_{p_{\perp}} / \sigma^{2}_{p} \approx 0.7\pm0.1$, irrespective of
the nature of the forcing. It is perhaps unfortunate that this situation arises; interestingly, the transverse
{\it velocity} power fraction is different for different forcing mechanisms, even at high Mach number (Federrath et al 2011).
We discuss the prospects of measuring the transverse velocity power fraction below.

\begin{figure}
\includegraphics[width=84mm]{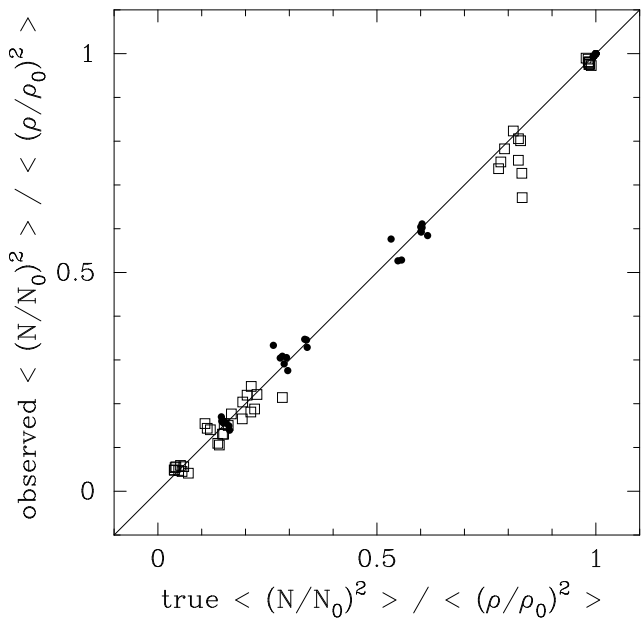}
\caption{Observed versus intrinsic values of the ratio $\langle (N/N_{0})^{2} \rangle / \langle (\rho/\rho_{0})^{2} \rangle$
measured in the numerical simulations. The ``observed'' values are calculated using the BFP method, which is accurate 
to about 10\% for statistically isotropic fields.}
\label{fig:ndr}
\end{figure}

Now we attempt to measure $\sigma^{2}_{p_{\perp}} / \sigma^{2}_{p}$ using only observationally-accessible
quantities. The analysis below is somewhat idealised, in that we assume certain auxiliary pieces of 
physical information are available, and we do not consider the effects of instrumental noise and beam smearing 
-- though we will make comments and recommendations at appropriate points.

\subsection{Normalised density variance}

First, we use the BFP method to estimate the 3D normalised density variance, $\sigma^{2}_{\rho/\rho_{0}}$,
using information contained solely in the observationally accessible normalised column density field,
namely $\sigma^{2}_{N/N_{0}}$ and the angular average of the column density power spectrum, $f(k)$. The
second term in equation~(\ref{allofit}) is then formed as: 
\begin{equation}
\frac{\langle (N/N_{0})^{2} \rangle}{\langle (\rho/\rho_{0})^{2} \rangle} = \frac{1+\sigma^{2}_{N/N_{0}}}{1+\sigma^{2}_{\rho/\rho_{0}}} .\nonumber
\end{equation}
In Figure~\ref{fig:ndr} we plot the observational value of this ratio versus the true quantity measured by
privileged access to the 3D density field. To make the data in this plot, we make use of all 3 possible orientations of the
simulation cubes to generate 3 column density fields per simulation. Note that large values of 
$\langle (N/N_{0})^{2} \rangle / \langle (\rho/\rho_{0})^{2} \rangle$ originate from the weakly-varying density
fields (low Mach number) while the smaller values originate from more variable density fields (higher Mach numbers) in which
small-scale structure suffers from a greater degree of line-of-sight averaging. In the observational context, such
measurements are subject to instrumental noise (resulting in a noise contribution to the column density variance and
column density power spectrum) and the effect of beam smearing of the telescope. These factors can be dealt with
using the methods outlined in BFP and Brunt (2010).

\begin{figure*}
\includegraphics[width=78mm,angle=-90]{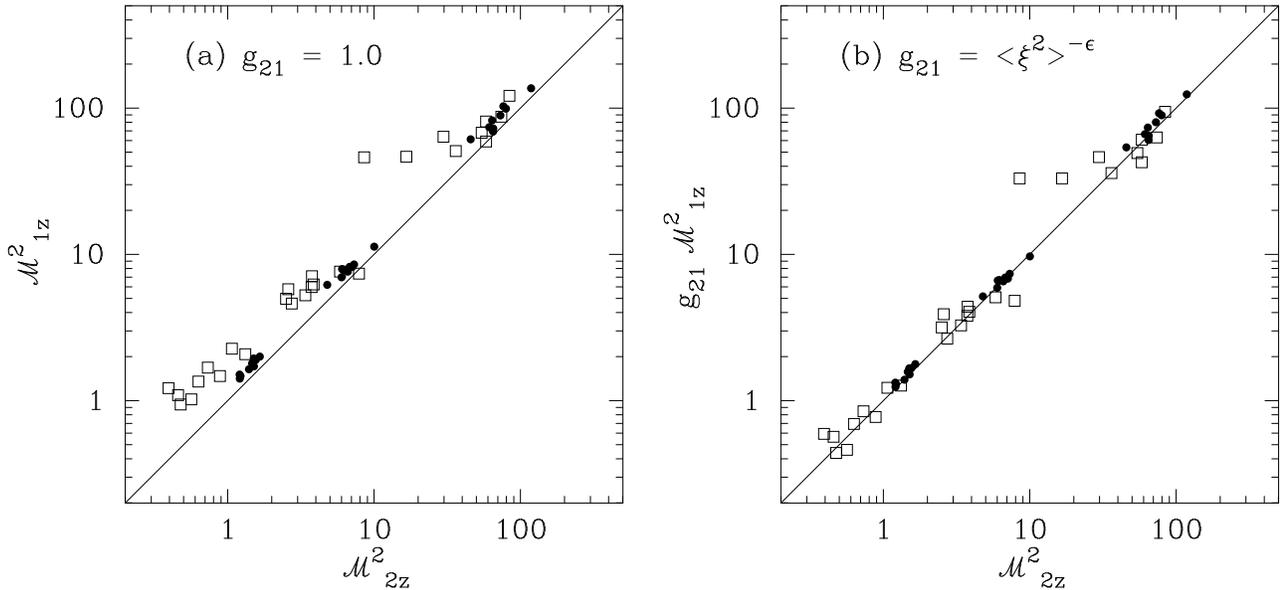}
\caption{(a) Estimate of ${\mathcal{M}}_{2z}^{2}$ naively assuming ${\mathcal{M}}_{2z}^{2} = {\mathcal{M}}_{1z}^{2}$; 
(b) Estimate of ${\mathcal{M}}_{2z}^{2}$ assuming ${\mathcal{M}}_{2z}^{2} = g_{21} {\mathcal{M}}_{1z}^{2}$.}
\label{fig:gcorr}
\end{figure*}

\begin{figure*}
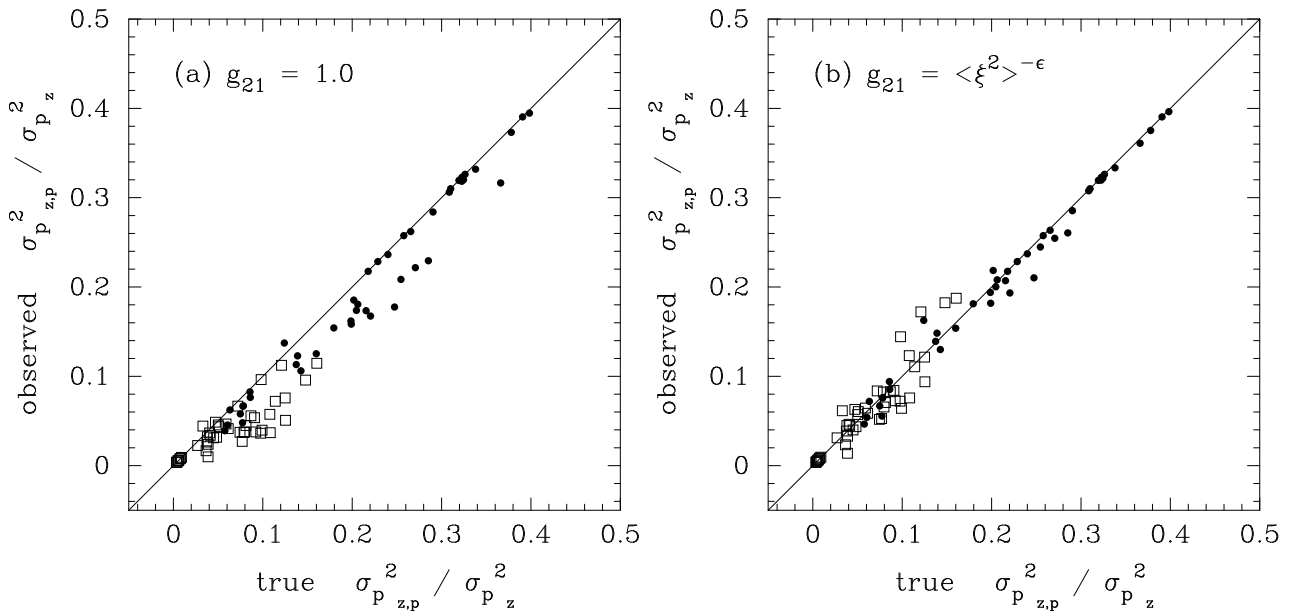

\includegraphics[width=84mm]{f6a.eps}
\includegraphics[width=84mm]{f6b.eps}
\caption{Observationally estimated values of the fraction of total $z$-momentum power projected into 2D assuming (a) $g_{21} = 1$ (i.e. no
statistical correction for density-velocity correlation) and (b) $g_{21} = \langle \xi^{2} \rangle^{-\epsilon}$, in both
cases plotted versus the true value of this fraction.}
\label{fig:projdata}
\end{figure*}

\subsection{Dispersion in $z$-axis momentum}

The next term in equation~(\ref{allofit}) we examine is the third term, $g_{21} \langle \rho v_{z}^{2} \rangle / \langle \rho \rangle$.
This term is designed to measure the total dispersion in $z$-axis momentum (divided by $\langle \rho^{2} \rangle$) as described
by equation~(\ref{usingg21}) -- or equivalently, the $z$-axis velocity dispersion calculated with a $\rho^{2}$ weight.
As mentioned previously, $\langle \rho v^{2} \rangle / \langle \rho \rangle$ can either be obtained from the
ratio $\langle W_{2} \rangle / \langle W_{0} \rangle$ or by the dispersion of the summed spectral line profile. The latter
option is probably best-suited to practical observational work, though the two are equivalent. Corrections for thermal/instrumental
broadening should be made. We point out here that for the low Mach number simulations (${\mathcal{M}} < 1$), the thermal broadening 
dominates over the turbulent motions of interest. Here we assume no thermal broadening (or exact accounting for it) which is
rather unrealistic. However, our aim here is to test the principle rather than the practice, since the low Mach number simulations
extend the range of $\sigma^{2}_{p_{\perp}} / \sigma^{2}_{p}$ available for analysis (see Figure~\ref{fig:tfrac}). We do not recommend that the 
method is applied practically to subsonic media, but expect instead that it will be applied to large molecular clouds where
the thermal broadening has a small influence on the dominant supersonic turbulent motions.

A further idealisation that we use here
is that we work with Mach numbers rather than velocity dispersions. This is just a convenient system in which to make comparisons between velocity
dispersions (in units of the squared sound speed), though we note that our estimation of the parameter $\epsilon$ needed to derive 
$g_{21}$ via equation~(\ref{epsiloninfinity}) requires the Mach number ${\mathcal{M}}_{1}$ to be known. In the observational context, this 
would require that an accurate measure of the sound speed is available. We have made a further assumption here, which is: the
use of equation~(\ref{epsiloninfinity}) assumes the hydrodynamic limit, since the fit was made to simulations that did not
include magnetic fields. There is an apparent dependence of $\epsilon$ on the Alfv\'{e}nic Mach number (see Figure~\ref{fig:evsma})
but this is only important in the limit ${\mathcal{M}}_{A} < 1$ and low sonic Mach number (${\mathcal{M}}_{1} \lesssim 5$).
We assume that auxiliary observations have been made to ensure that these conditions are not met, though the consequences
of erroneously assuming they are not met when in fact they are (at least within the range of conditions covered by the 
above analysis) are not severe. In the low ${\mathcal{M}}_{A}$, low ${\mathcal{M}}_{1}$ regime, $\epsilon$ is slightly 
smaller than that which holds in the hydrodynamic limit ($\epsilon_{\infty}$), and therefore the true correction factor 
lies between unity and $g_{21}(\epsilon_{\infty})$. As shown below, the practical difference between these two cases 
is tolerably small. 

From the simulation data, we can directly extract the quantity 
${\mathcal{M}}_{1z} = \langle \rho v^{2}_{z} \rangle / \langle \rho \rangle c^{2}_{s} $, 
which is the $\rho$-weighted $z$-velocity dispersion in units of the squared sound speed. 
The correction factor $g_{21}$ is designed to convert this into
the $\rho^{2}$-weighted $z$-axis Mach number: ${\mathcal{M}}_{2z} = \langle \rho^{2} v^{2}_{z} \rangle / \langle \rho^{2} \rangle c^{2}_{s}$. 
To calculate $\epsilon$ we assume isotropy and take ${\mathcal{M}}_{1} = \sqrt{3} {\mathcal{M}}_{1z}$,
allowing us to calculate $g_{21} = \langle (\rho/\rho_{0})^{2} \rangle^{-\epsilon}$
using equation~(\ref{eq:g21}) with $\langle (\rho/\rho_{0})^{2} \rangle$ already calculated above. In Figure~\ref{fig:gcorr}(a)
we plot $M^{2}_{1z}$ versus $M^{2}_{2z}$ (measured with privileged access) for the supersonic data, to illustrate the 
consequences of the naive assumption that $M^{2}_{1z}$ is a good surrogate for $M^{2}_{2z}$. Figure~\ref{fig:gcorr}(b)
shows the effect of assuming that $M^{2}_{2z} = g_{21} M^{2}_{1z}$, which is rather better, though there are a couple of outliers
that resist correction (these are notably anisotropic fields). In Figure~\ref{fig:gcorr} we have again used all 3 available
choices for the $z$-axis in each simulation. In the analysis below, we will document the effect of including or ignoring
the correction factor $g_{21}$.

\begin{figure*}
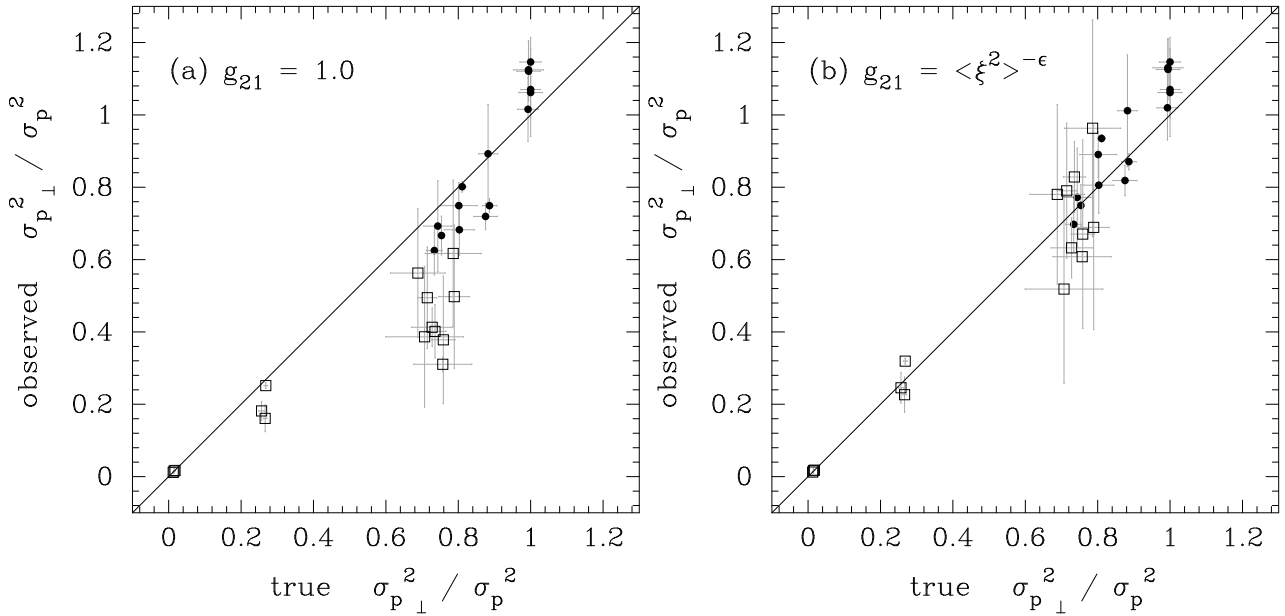

\includegraphics[width=84mm]{f7a.eps}
\includegraphics[width=84mm]{f7b.eps}
\caption{Observationally estimated values of the fraction of 3D momentum power in transverse modes assuming (a) $g_{21} = 1$ (i.e. no
statistical correction for density-velocity correlation) and (b) $g_{21} = \langle \xi^{2} \rangle^{-\epsilon}$, in both
cases plotted versus the true value of this fraction.}
\label{fig:finalplot2}
\end{figure*}

\subsection{Projected solenoidal fractions}

The first factor in equation~(\ref{allofit}) is trivially computed using the observationally accessible 
$\langle W^{2}_{1} \rangle / \langle W^{2}_{0} \rangle$. We have already mentioned accounting for the
contribution of noise variance to $\langle W^{2}_{0} \rangle$, and it remains to deal with the role of
noise in calculating $\langle W^{2}_{1} \rangle$, which is a bit more complicated. We just mention here that
prescriptions for accounting for noise in centroid velocity measurements are available (e.g. Kleiner \& Dickman 1985;
Miesch \& Bally 1994; Brunt \& Mac Low 2004) and a suitable modification of these should be made. 

Now with all three terms in equation~(\ref{allofit}) assembled,
we can construct an observationally-accessible estimate of $\sigma^{2}_{p_{z,p}} / \sigma^{2}_{p_{z}}$ 
(i.e. the fraction of $z$-momentum power projected into 2D) for comparison with the same ``true'' quantity directly 
extracted from the data with privileged access. Figure~\ref{fig:projdata} shows this comparison, (a) without and
(b) with the correction factor $g_{21}$ applied. In general, equation~(\ref{allofit}) provides an accurate
observational measure of the ratio $\sigma^{2}_{p_{z,p}} / \sigma^{2}_{p_{z}}$.

Note that there are two main effects
that control this ratio. First, if the fraction of compressive (longitudinal) momentum modes is high, then the
ratio $\sigma^{2}_{p_{z,p}} / \sigma^{2}_{p_{z}}$ will be low since only transverse modes are projected into 2D.
This is (partially) why the compressively-forced simulations display small values of $\sigma^{2}_{p_{z,p}} / \sigma^{2}_{p_{z}}$.
Second, a higher density variance will typically result in a lower fraction of momentum variance projected into
2D since such fields suffer more averaging/smoothing of small-scale structure (e.g. as mentioned above in the
discsussion of Figure~\ref{fig:ndr}). This means that even though the transverse momentum fractions are approximately
the same at high Mach numbers regardless of the forcing mechanism (see Figure~\ref{fig:tfrac}) the compressive forcing
drives higher density variance (Federrath et al 2008) which results in a greater degree of line-of-sight averaging of
small-scale structure, and therefore lower values of $\sigma^{2}_{p_{z,p}} / \sigma^{2}_{p_{z}}$.
The highest values of $\sigma^{2}_{p_{z,p}} / \sigma^{2}_{p_{z}}$ reached ($\sim 1/3$) originate from the
subsonic solenoidally-forced simulations which have very nearly uniform density. In this case, trivially,
$\sim$~1/3 of the momentum variance is projected along one of the three spatial axes.

\subsection{Deprojection}

The final step in the analysis is the de-projection via equation~(\ref{allofit2}) to estimate the
fraction of momentum power in transverse modes in 3D 
(assuming that $\sigma^{2}_{p_{\perp}} / \sigma^{2}_{p} \approx \sigma^{2}_{p_{z\perp}} / \sigma^{2}_{p_{z}}$). 
The input to the de-projection factor is 
just $f_{\perp}(k)$, obtained from the angular average of the power spectrum of $W_{1}$. In an observational
context this should be corrected for noise (using a suitable modification of the methods in 
Brunt \& Mac Low 2004 or Brunt 2010) and treatment of
the effect of the telescope beam pattern should be included (see BFP for a discussion of this).

In Figure~\ref{fig:finalplot2} we compare the observationally derived values
of $\sigma^{2}_{p_{\perp}} / \sigma^{2}_{p}$ to those measured exactly with privileged access to
the momentum field in 3D. Figure~\ref{fig:finalplot2}(a) shows the comparison with no statistical
correction for density-velocity corrlation, while Figure~\ref{fig:finalplot2}(b) shows the results with
the $g_{21}$ correction applied (we have applied the correction to all data, not just the supersonic fields).
In the plots, the plotted points are the mean values of $\sigma^{2}_{p_{\perp}} / \sigma^{2}_{p}$
obtained by averaging the results over all 3 spatial axes, while the error bars represent the
standard deviation about the mean. This is to show the recovery of $\sigma^{2}_{p_{\perp}} / \sigma^{2}_{p}$
from the same field seen from different orientations. Overall, the observational recovery of the intrinsic 
$\sigma^{2}_{p_{\perp}} / \sigma^{2}_{p}$ values is good, albeit with relatively large scatter for the
high Mach number compressively-forced simulations. 

Though the solenoidally-forced simulations
alone span a limited range in $\sigma^{2}_{p_{\perp}} / \sigma^{2}_{p}$, the recovery is reliable, albeit with
a slight overestimation for intrinsic $\sigma^{2}_{p_{\perp}} / \sigma^{2}_{p} \approx 1$ (applicable to the
subsonic, $\sim$~uniform density fields). If the turbulence was strongly magnetized (i.e.
$\mathcal{M}_{s} \lesssim 5$, $\mathcal{M}_{A} \lesssim 1$), but assumed not to be,
then the $g_{21}$ factors derived from $\epsilon_{\infty}$ would lead to slighty over-estimated solenoidal
fractions, though for the range of Mach numbers studied here, this is only at the $\sim$~10\% level. This
should be tested directly in future studies, where the (probably) more important question of anistropy 
should be assessed. 

\section{Discussion}

While the simulated fields here contain a number of idealisations (Fourier-space driving, periodic boundary
conditions, lack of self-gravity, strict isothermality) the above analysis has demonstrated that, in
principle, the fraction of momentum power in transverse (solenoidal) modes, $\sigma^{2}_{p_{\perp}} / \sigma^{2}_{p}$,
may be measureable from observations. (We have outlined practical observational considerations at appropriate points above.)
For the subsonic fields, to extend the range of intrinsic $\sigma^{2}_{p_{\perp}} / \sigma^{2}_{p}$ available,
we have assumed (unrealistically) that the dominant thermal broadening terms have been accounted for.

We envision that the above method is 
best-suited to the study of nearby supersonically-turbulent molecular clouds for which high sensitivity, 
high spatial dynamic range spectral line data are available. So, how relevant
are the simulated supersonic fields to typical conditions met in molecular clouds that (currently) could be 
analysed with the above model? We argue that the key condition that must be satisfied is that of statistical
isotropy, rather than any shortcomings in the physical simulations -- i.e. the simulations simply create an intrinsic 
ratio $\sigma^{2}_{p_{\perp}} / \sigma^{2}_{p}$ that we can set out to measure. If strong magnetic fields are present, then
this could cause significant anisotropy in the density and velocity fields that we have not accounted for here. 
The assumption of isothermality may be slightly problematic, since the method relies (weakly) on applying corrections for thermal broadening and 
deriving the Mach number (to estimate $g_{21}$). The assumption of a single Mach number is also problematic
if the method is to be applied to the atomic ISM. It is also possible of course that the detailed physics may affect the
statistical correction factors $g_{21}$, a point to which we will return below. 

\subsection{Anisotropies and discontinuities}

The next steps in the 
testing of the method should focus on possible complications and biases induced by anisotropies.
In fact, there are suggestions that compressive modes may be suppressed in highly anistropic media 
(Hansen, McKee, \& Klein 2011),
which would be very interesting to test observationally once the practicalities of doing so are understood more fully.
Testing (or extension) of the method to MHD case with different compressive/solenoidal fractions is also worth pursuing. 
Filamentary structure, by itself, poses no
particular problem as long as the filament orientations are statistically isotropic. In practice, good evidence for
isotropy in projected 2D is needed, as well as some confirmation that the cloud's line-of-sight depth is comparable
to its projected extent. In the field selection, one must also pay attention to the boundary conditions. In our
simulations, these are periodic, so there are no problems arising from edge discontinuities. Fields for analysis should
be selected so that no significant edge discontinuities exist, though we are aided in this selection by the $\rho$-
(or $N$-) weighting of the relevant fields so that (to the extent that any ``cloud'' is truly isolated) a suitable
region may be defined where $W_{0}$, $W_{1}$, and $W_{2}$ are sufficiently close to zero that this condition is
satisfied. Edge-tapering (e.g. Brunt \& Mac Low 2004) or padding (e.g. Brunt 2010) may be applied to 
sufficiently large fields, with negligible quantitative consequences. 

\begin{figure}
\includegraphics[width=84mm]{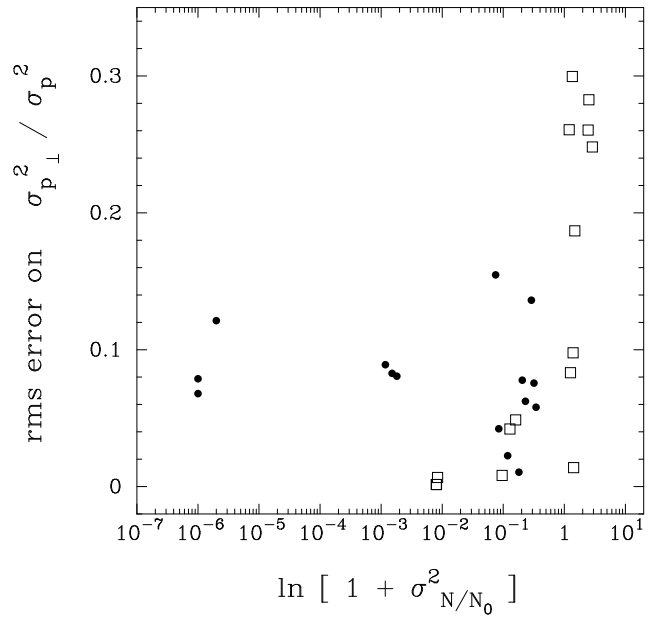}
\caption{The root mean square uncertainty on the measured $\sigma^{2}_{p_{\perp}} / \sigma^{2}_{p}$ ratios from Figure~\ref{fig:finalplot2},
plotted against a measure of the column density variability, $\ln{\left(1 + \sigma^{2}_{N/N_{0}}\right)}$.}
\label{fig:rmserror}
\end{figure}

\subsection{Variability}

In terms of the reliability of the recovered $\sigma^{2}_{p_{\perp}} / \sigma^{2}_{p}$, of most concern is the relatively
high degree of scatter in the compressively-forced fields. The scatter is mainly a consequence of the high degree of
variability in these fields, so that variances are contributed to strongly by a small number of extreme field values,
perhaps magnifying any anisotropic effects. To quantify this, in Figure~\ref{fig:rmserror} we plot the
root mean square uncertainty on the measured $\sigma^{2}_{p_{\perp}} / \sigma^{2}_{p}$ ratios 
against a measure of the column density variability, $\ln{\left(1 + \sigma^{2}_{N/N_{0}}\right)}$ (which would
be the logarithmic variance, $\sigma_{ln(N/N_{0})}$, in the case of a lognormal PDF). 
Figure~\ref{fig:rmserror} shows that typical errors
are at the $\sim$~10\% level or better for $\ln{\left(1 + \sigma^{2}_{N/N_{0}}\right)} \lesssim 1$, but then increase
sharply for $\ln{\left(1 + \sigma^{2}_{N/N_{0}}\right)} \gtrsim 1$.

\subsection{Variations in solenoidal/compressive fractions}

It is evident from Figure~\ref{fig:tfrac} that the $\sigma^{2}_{p_{\perp}} / \sigma^{2}_{p}$ ratio does
not allow observational discrimination between solenoidal and compressive forcing, since the 
$\sigma^{2}_{p_{\perp}} / \sigma^{2}_{p}$ ratios appear to converge to $\sim$~3/4 at high Mach numbers
independent of the nature of the forcing
(it is probably unlikely that these curves cross, though we cannot say more at present). However,
given that observational estimates of $\sigma^{2}_{p_{\perp}} / \sigma^{2}_{p}$ have not yet been made,
it would be an interesting and important test to see if this ratio ($\sim$~0.7 $\pm$ 0.1) is realised in
nature and whether there may be systematic variations in different environments. Note that, though
mathematically well-defined, the forcing scheme is somewhat idealised physically -- assigning accelerations 
in Fourier space, which generate non-local accelerations in direct space. It is possible that
the $\sigma^{2}_{p_{\perp}} / \sigma^{2}_{p}$ ratios seen here may not necessarily be replicated
in real molecular clouds, where large-scale driving sources could include supernovae or spiral shocks.

One could also consider the initially point-like energy injections from outflows in
localised regions generating different $\sigma^{2}_{p_{\perp}} / \sigma^{2}_{p}$ ratios. 
It may also be possible that $\sigma^{2}_{p_{\perp}} / \sigma^{2}_{p}$ could evolve with time in 
decaying conditions (our simulations here are continually driven) or due to an increasing importance
of self-gravity over time. Observational application of this
method therefore may be more revealing and interesting than simply confirming the ``default'' $\sim 0.7 \pm 0.1$ 
ratio seen in the current simulations. A framework for investigating the
respective roles of solenoidal and compressive forcing in determining the normalised density 
variance has been recently presented by Konstandin et al (2012b), 
and it would be interesting to compare (or combine) that method with ours.

The inclusion of self-gravity will be particularly interesting. Federrath et al (2011) have
shown that gravitational collapse produces a high fraction of longitudinal modes, which are
later converted to solenoidal modes. It may therefore be possible to search for this signature
in molecular clouds, after we have validated the method for the self-gravitating case. This is beyond
the scope of the current work, but we note that the geometrical constraints, that lead to projection
of solenoidal modes only, will not be affected by inclusion of self-gravity. It is true, however, that
self-gravity will have a notable effect on the density PDF, and presumably on the degree of density--velocity
correlation. A number of theoretical studies have looked at the generation of $\sim$power--law tails in
density PDFs (e.g. Klessen 2000; Kritsuk, Norman, \& Wagner 2011; Federrath \& Klessen 2013; Girichidis 2014)
and corresponding features have been seen observationally: early-time PDFs appear $\sim$lognormal in form,
while more evolved clouds have more prevalent power--law tails in their density PDFs (e.g. Kainulainen et al 2009;
Schneider et al 2013; Schneider et al 2014; Kainulainen et al 2014).

\subsection{Improvements}

In the further testing of this model (in addition to the key question of anisotropy) there are two
main areas that should be investigated. First, in terms of numerics, we need more information on the
role of density-velocity correlations (Section~4) by measuring $\epsilon$ over a greater range of
physical conditions, including the effects of self-gravity, as mentioned above, 
and a more detailed investigation of magnetic fields. 
In the analysis completed above, we have introduced the concept of $\epsilon$
and the correction factors $g_{mn}$ and demonstrated their use in practice. Admittedly, there was
a degree of internal ``tuning'' of $\epsilon$ employed, though we did support the measured $\epsilon$ values
with independent data, and the correction terms $g_{21}$ here
are close enough to unity to be of not overwhelming concern. More important, perhaps, is the
possible role that the statistical system presented in Section~4 may play in other contexts for
understanding density-velocity correlations in turbulent media. 

The second main avenue for improvement will be in the investigation of more mundane observational considerations, such as
excitation and opacity. The analysis presented above was completed under the assumption of
uniform excitation and in the optically-thin limit, and therefore serves as a baseline for
quantifying how variable excitation and finite opacity affect the method. Naively, this
would make the measurements of $\sigma^{2}_{p_{\perp}} / \sigma^{2}_{p}$ less reliable, and
possibly biased, but may also have the advantageous affect of taming some of the field variability.

\subsection{Other moments}

Finally, we have based the method around the recovery of the fraction of {\it momentum} power, 
$\langle {\rho^{2}}v^{2} \rangle$, in transverse modes simply because the momentum field satisfies the observational requirement
that it is projected into 2D unweighted by any other physical fields. One could also consider 
the fraction of velocity dispersion, $\langle v^{2} \rangle$, in transverse modes (which is sensitive
to the forcing mechanism -- Federrath et al 2011) or indeed the 
fraction of energy, $\langle (1/2){\rho}v^{2} \rangle$,
in transverse modes -- either of which, arguably, more naturally spring to mind as relevant
quantities to measure. The prospects for measuring either of the two above-named alternatives to
momentum power are rather dim however, since neither field is accessible as a projected quantity
observationally (except in the case of uniform density when all three definitions 
are equivalent). 

In the general case where the density is variable,
neither the velocity nor energy can be isolated for independent study. For example, note that the 
$z$-component of the momentum field, which is projected unweighted into 2D, is the product 
of the density, $\rho$, and $z$-velocity, $v_{z}$. 
Therefore the crucially-important (projected) power spectrum of $W_{1}$ involves the 3D {\it convolution} 
of the Fourier transforms of $\rho$ and $v_{z}$, of which a single plane ($k_{z} = 0$) is available for
analysis. To isolate the velocity contribution, one could in principle imagine deconvolving the 
density contribution for a simply-structured density field (e.g. a 3D Gaussian or spatial power-law), but for 
the highly complex and variable density fields encountered in the ISM, this appears futile. One possible 
way to proceed is to form an understanding of the relation between
the three quantities ($\langle v^{2} \rangle$, $\langle (1/2){\rho}v^{2} \rangle$, and $\langle {\rho^{2}}v^{2} \rangle$)  
by a means similar to the statistical system presented in the Appendix -- because the different quantities are 
just different $\rho^{q}$-weighted velocity dispersions.
The stumbling block is simply that the Fourier space power distributions may be rather different, so that
the deprojection factors become unreliable. In this case, an understanding of any scale-dependency of the 
statistical relationship between $\langle v^{2} \rangle$, $\langle (1/2){\rho}v^{2} \rangle$, 
and $\langle {\rho^{2}}v^{2} \rangle$ would be required, as well as an understanding of their transverse/longitudinal
mode dependence.
These considerations are beyond the scope of the current paper, but may be a worthwile pursuit in future studies.

\section{Summary}

In this paper we have introduced an observational method for measuring the
fraction of momentum power in solenoidal modes in a turbulent cloud, and confirmed
its applicability using hydrodynamic numerical simulations. The method
is best-suited to application in nearby molecular clouds for which high sensitivity,
high spatial dynamic range spectral line observations are available. The principal
limitation of the method at present is its reliance on the assumption of isotropy.
Further work is needed to examine the impact of anisotropy imposed by (e.g.) magnetic
fields (BFP) or anisotropic driving of turbulence (Hansen et al 2011). Isotropy aside,
the main limiting factor in the accuracy of the model is variability in the physical
fields (density, momentum). 
We have also introduced a statistical framework for describing density-velocity correlations
in turbulent media that should be of relevance beyond its application here. 

\section*{Acknowledgments}

A big thanks to Daniel Price for allowing us use of the auxiliary numerical simulations, to
Maria Cunningham for allowing us access to the Delta Quadrant Survey data, and
to Dave Acreman for much-needed help with Figures 1 and 2.
C.~B. is funded in part by the UK Science and Technology Facilities Council grant ST/J001627/1 
(``From Molecular Clouds to Exoplanets'') and the ERC grant ERC-2011-StG\_20101014 (``LOCALSTAR''),
both held at the University of Exeter. 
C.~F.~acknowledges funding provided by the Australian Research Council under the Discovery 
Projects scheme (grant DP110102191). Supercomputing time at the Leibniz Rechenzentrum 
(project pr32lo) and at the Forschungszentrum J\"ulich (project hhd20) are gratefully acknowledged. 
The software used in this work was in part developed by the DOE-supported ASC / Alliance Center for 
Astrophysical Thermonuclear Flashes at the University of Chicago.

\appendix{}
\section{Density--Velocity Correlations}{\label{sec:epicocity}}

In Section~2.3. we introduced a correction factor $g_{21}$, needed to convert a 
$\rho$-weighted velocity dispersion into a $\rho^{2}$-weighted velocity dispersion. 
Here we discuss the procedural steps necessary to evaluate the relationship between 
$\langle \rho v^{2}_{z} \rangle/\langle \rho \rangle$ and $\langle \rho^{2} v^{2}_{z} \rangle/\langle \rho^{2} \rangle$,
in order to derive the statistical correction factor $g_{21}$ introduced in Section~2.3.

As we demonstrate below, if the density and velocity field are statistically uncorrelated (i.e. they have independent
probability distributions) then 
$\langle \rho v^{2}_{z} \rangle/\langle \rho \rangle = \langle \rho^{2} v^{2}_{z} \rangle/\langle \rho^{2} \rangle$,
and therefore $g_{21} = 1$. In general
this will not be the case, so in the following we introduce and test a simple method for quantifying the
degree of density-velocity correlation and establish a simple prescription for converting between
different $\rho^{q}$-weighted velocity dispersions. After theoretical development (Setions A1, A2) and testing (Section A3),
we examine existing observational constraints on these corrections (Section A4).

\begin{figure*}
\includegraphics[width=104mm,angle=-90]{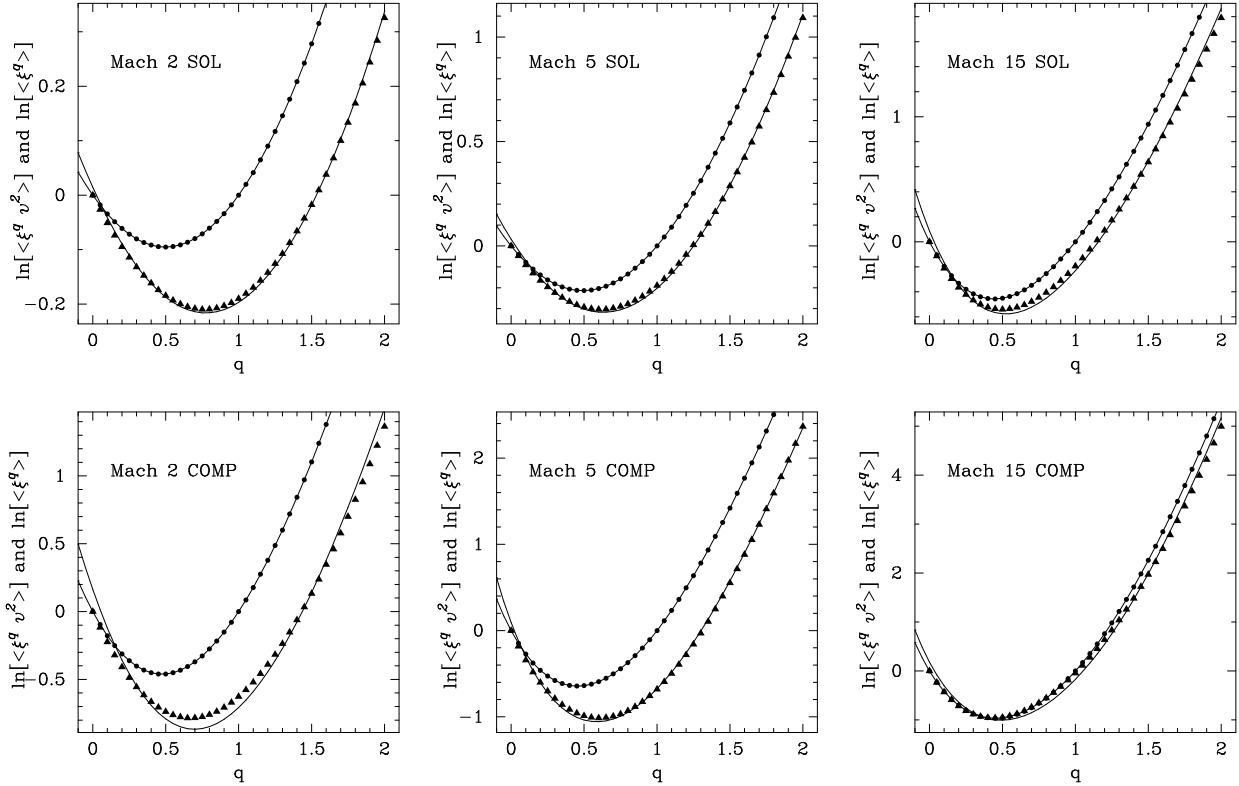}
\caption{The spectrum of moments for $\langle \xi^{q} \rangle$ (dots) and $\langle \xi^{q} v^{2} \rangle$ (triangles) versus $q$. The curve (described in
the text) is fitted to $\langle \xi^{q} \rangle$ and shifted to best match $\langle \xi^{q}v^{2} \rangle$ in order to determine $\epsilon$.}
\label{fig:epicplot}
\end{figure*}

\subsection{Theoretical Development}

The velocity dispersion, $\sigma^{2}_{q}$, calculated using a $\rho^{q}$ weighting can be written as:
\begin{eqnarray}
\sigma^{2}_{q} \lefteqn{= \frac{\frac{1}{V} \displaystyle\int_{V} {\mathrm{d}}V \;  \rho^{q} v^{2}}{\frac{1}{V} \displaystyle\int_{V} {\mathrm{d}}V \; \rho^{q}} = \frac{\langle \rho^{q} v^{2}\rangle}{\langle \rho^{q} \rangle} } \nonumber \\ 
\lefteqn{= \frac{\frac{1}{V} \displaystyle\int_{V} {\mathrm{d}}V \;  \xi^{q} v^{2}}{\frac{1}{V} \displaystyle\int_{V} {\mathrm{d}}V \; \xi^{q}} = \frac{\langle \xi^{q} v^{2}\rangle}{\langle \xi^{q} \rangle} ,} 
\label{eq:ep1}
\end{eqnarray}
where {\bf $\rho_{0} = \langle \rho \rangle$,} $\xi = \rho/\rho_{0}$, $V$ is the volume containing the fields and angle brackets denote
spatial averages. In what follows, we use the more convenient variable $\xi$ to perform calculations.

The volumetric integration can be replaced by integrals over $\xi$ and $v$ as follows:
\begin{equation}
\sigma^{2}_{q} = \frac{\displaystyle\int_{0}^{\infty}\displaystyle\int_{-\infty}^{\infty} {\mathrm{d}}\xi \; {\mathrm{d}}v\; P_{\xi}(\xi) P_{v}(v) \; \xi^{q} v^{2}}{\displaystyle\int_{0}^{\infty} {\mathrm{d}}\xi \; P_{\xi}(\xi) \; \xi^{q}} ,
\end{equation}
where $P_{\xi}(\xi)$ and $P_{v}(v)$ are the probability distribution functions (PDFs) of $\xi$ and $v$ respectively. To account for correlations
between density and velocity, we consider the case now
where $P_{v}(v)$ is an implicit function of $\xi$, so that completing the $v$ integral leads to:
\begin{equation}
\sigma^{2}_{q} = \frac{\displaystyle\int_{0}^{\infty} {\mathrm{d}}\xi \; P_{\xi}(\xi) \; \xi^{q} \sigma^{2}_{v}(\xi) }{\displaystyle\int_{0}^{\infty} {\mathrm{d}}\xi \; P_{\xi}(\xi) \; \xi^{q}} ,
\label{eq:sigvx}
\end{equation}
where $\sigma^{2}_{v}(\xi)$ is a density-dependent velocity dispersion. We propose a simple form
for this dispersion as follows:
\begin{equation}
\sigma^{2}_{v}(\xi) = h(\xi) \sigma^{2}_{00} ,
\label{vdispeq1}
\end{equation}
where $\sigma^{2}_{00}$ is a constant, and further propose that:
\begin{equation}
h(\xi) = \xi^{-\epsilon}  ,
\label{vdispeq2}
\end{equation}
with the expectation (but not requirement) that $\epsilon$ is a small positive constant, 
so that higher densities are associated with smaller velocity dispersions. 
Equation~(\ref{eq:sigvx}) then becomes:
\begin{equation}
\sigma^{2}_{q} = \frac{\displaystyle\int_{0}^{\infty} {\mathrm{d}}\xi \; P_{\xi}(\xi) \; \xi^{q-\epsilon} \sigma^{2}_{00} }{\displaystyle\int_{0}^{\infty} {\mathrm{d}}\xi \; P_{\xi}(\xi) \; \xi^{q}} ,
\label{eq:sigvxa}
\end{equation}
with the immediate result:
\begin{equation}
\sigma^{2}_{q} =   \frac{\langle \xi^{q-\epsilon} \rangle \sigma^{2}_{0} }{ \langle \xi^{q} \rangle \langle \xi^{-\epsilon} \rangle} ,
\label{eq:sigvxb}
\end{equation}
where we have written the unweighted ($q = 0$) velocity dispersion, $\sigma^{2}_{0}$, as:
\begin{equation}
\sigma^{2}_{0} = \sigma^{2}_{00} \langle \xi^{-\epsilon} \rangle .
\end{equation}
Note that if density and velocity are statistically uncorrelated ($\epsilon = 0$) then
Equation~(\ref{eq:sigvxb}) just gives $\sigma^{2}_{q} = \sigma^{2}_{0}$ for all $q$.

\begin{figure}
\includegraphics[width=84mm]{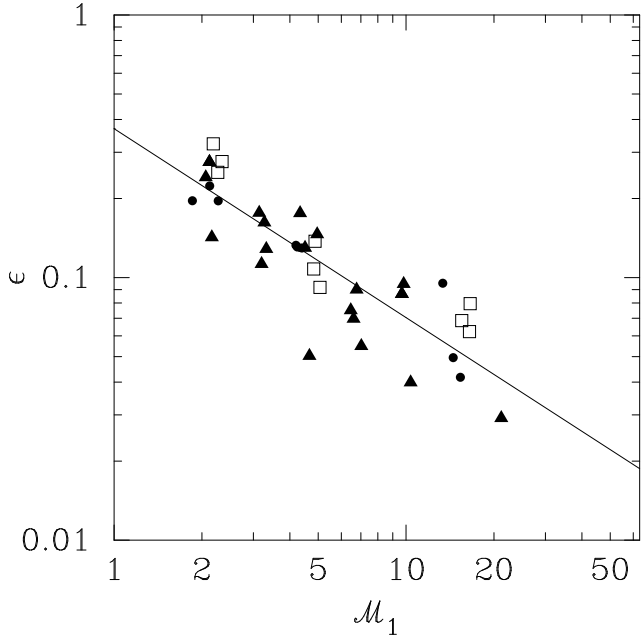}
\caption{Plot of $\epsilon$ versus 3D density-weighted Mach number, ${\mathcal{M}}_{1}$, derived from the hydrodynamic grid
simulations (dots: solenoidal forcing; open squares: compressive forcing; triangles: hydrodynamic (SPH) simulations with
solenoidal forcing from BFP.}  
\label{fig:epichydro}
\end{figure}

Using Equations~(\ref{eq:ep1}) and (\ref{eq:sigvxb}) without the normalising denominator, we may also write:
\begin{equation}
\langle \xi^{q} v^{2} \rangle  =   \frac{\langle \xi^{q-\epsilon} \rangle \sigma^{2}_{0} }{ \langle \xi^{-\epsilon} \rangle} ,
\label{eq:sigvxc}
\end{equation}
i.e. that the spectrum of moments $\langle \xi^{q} v^{2}\rangle (q)$ is a scaled, shifted version of the spectrum
of moments  $\langle \xi^{q} \rangle (q)$. 

\subsection{Formulation for Lognormal Density PDFs}

Knowledge of $\langle \xi^{q} v^{2}\rangle/\langle \xi^{q} \rangle$ would be very useful, but it in turn
requires knowledge of $\langle \xi^{q} \rangle (q)$ which itself is inaccessible
observationally. However, progress can be made by employing an analytic form for the PDF of normalised
density, $P_{\xi}(\xi)$. For isothermal turbulence, a lognormal PDF for $\xi$ is a reasonable
approximation (V\'{a}zquez-Semadeni 1994; Padoan, Nordlund, \& Jones 1997; Federrath et al 2008; Kainulainen et al 2009),
for which:
\begin{equation}
\langle \xi^{q} \rangle = \exp \left[ q \langle \ln(\xi) \rangle + \frac{1}{2} q^{2} \sigma^{2}_{\ln(\xi)} \right] .
\end{equation}
The normalisation of the field to $\langle \xi \rangle = 1$ requires that $\langle \ln(\xi) \rangle = -\frac{1}{2} \sigma^{2}_{\ln(\xi)}$,
so that
\begin{equation}
\langle \xi^{q} \rangle = \exp \left[ \frac{1}{2} \sigma^{2}_{\ln(\xi)} (q^{2} - q) \right] = \langle \xi^{2} \rangle^{\frac{1}{2}(q^{2} - q)},
\end{equation}
where we have made use of $\sigma^{2}_{\ln(\xi)} = \ln(1 + \sigma^{2}_{\xi}) = \ln(\langle \xi^{2} \rangle)$ in the last step.
The predicted form for $\langle \xi^{q} \rangle$ (in the case of a lognormal density PDF) agrees well with the
moment spectra shown in Figure~\ref{fig:epicplot}: i.e. $\ln{(\langle \xi^{q} \rangle)}$ is parabolic against $q$, equal to unity at
$q = 0$ and $q = 1$, and reaches a minimum at $q = 0.5$.

Inserting this result into Equation~(\ref{eq:sigvxc}) then gives:
\begin{equation}
\langle \xi^{q} v^{2} \rangle = \langle \xi^{2} \rangle^{\frac{1}{2}(q^{2} - q - 2q\epsilon)} \sigma^{2}_{0},
\end{equation}
and Equation~(\ref{eq:sigvxb}) becomes:
\begin{equation}
\sigma^{2}_{q} = \frac{\langle \xi^{q} v^{2} \rangle}{\langle \xi^{q} \rangle}  = \langle \xi^{2} \rangle^{-q\epsilon} \sigma^{2}_{0} .
\end{equation}
We define velocity dispersion ratios, $g_{mn}$, as follows:
\begin{equation}
g_{mn} = \frac{\sigma^{2}_{m}}{\sigma^{2}_{n}} = \frac{\langle \rho^{m} v^{2} \rangle/\langle \rho^{m} \rangle}{\langle \rho^{n} v^{2} \rangle/\langle \rho^{n} \rangle} = \langle \xi^{2} \rangle^{-(m-n)\epsilon}  .
\label{eq:vdisprat}
\end{equation}
In the next Section we will make use of the following ratio:
\begin{equation}
g_{21} = \frac{\sigma^{2}_{2}}{\sigma^{2}_{1}} = \frac{\langle \rho^{2} v^{2} \rangle/\langle \rho^{2} \rangle}{\langle \rho v^{2} \rangle/\langle \rho \rangle} = \langle \xi^{2} \rangle^{-\epsilon} .
\label{eq:g21}
\end{equation}
Note that if the density field is uniform ($\langle \xi^{2} \rangle = 1$) then $g_{21} = 1$. In
general, the assumption $\epsilon = 0$ requires that there be no statistical correlation
between density and velocity. To apply this correction observationally, a measurement of $\langle \xi^{2} \rangle$
is required. This is can be done, since the 3D normalised density dispersion, $\sigma^{2}_{\xi}$, and
therefore $\langle \xi^{2} \rangle = 1 + \sigma^{2}_{\xi}$
can be estimated by the BFP method.

\subsection{Numerical Testing}

The applicability of Equation~(\ref{eq:sigvxc}) to the numerical
density and velocity fields can be tested by calculating and comparing $\langle \xi^{q} v^{2}\rangle (q)$
and $\langle \xi^{q} \rangle (q)$. From the simulated data, moments between $q = 0$ and $q = 2$ in steps of 0.05
were calculated in the zero momentum frame, using the full 3D velocity field. 
To ensure that the resulting values of $\epsilon$ are
not simply ``tuned'' to the numerical simulations analysed in this paper, we use a much larger sample
of density and velocity fields by including the simulations from BFP. These include both hydrodynamic simulations
(using smoothed particle hydrodynamics) and MHD simulations (using grid calculations) -- see Section~3 and BFP for
more details. We only consider supersonic fields in this analysis, since the influence of density
fluctuations in subsonic fields is minimal -- we will return to this point at the end of this Section.

After calculation of the moments, a 6$^{th}$-order polynomial
in $q$ ($\sum_{n=0}^{6} c_{n} q^{n}$) was fitted to $\ln{(\langle \xi^{q} \rangle)}$. Following this, the same polynomial with 
the same (fixed) $c_{n}$ was fitted to $\ln{(\langle \xi^{q} v^{2}\rangle)}$ with additional offset ($\epsilon$) and scaling factor ($A$) -- 
i.e. $ \ln(A) + \sum_{n=0}^{6} c_{n} (q - \epsilon)^{n}$.

\begin{figure}
\includegraphics[width=84mm]{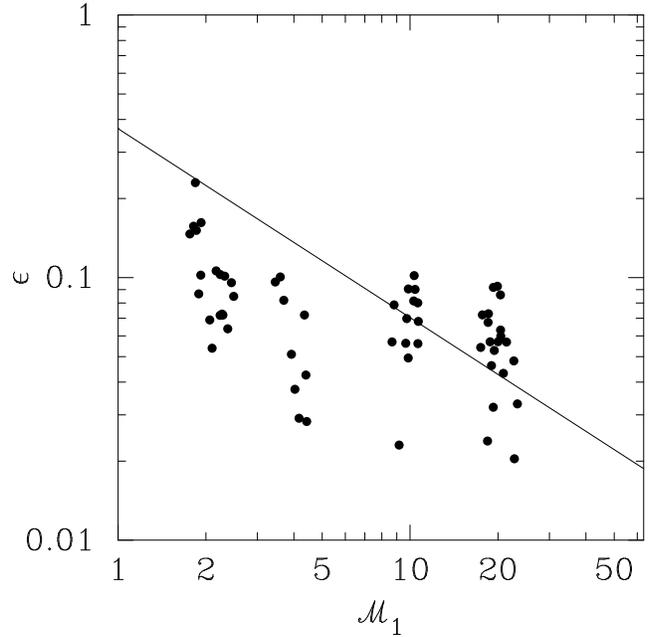}
\caption{Plot of $\epsilon$ versus 3D density-weighted Mach number, ${\mathcal{M}}_{1}$, derived from the MHD simulations (BFP). For reference
we have shown the fitted line from Figure~\ref{fig:epichydro}.}
\label{fig:epicmhd}
\end{figure}

\begin{figure}
\includegraphics[width=84mm]{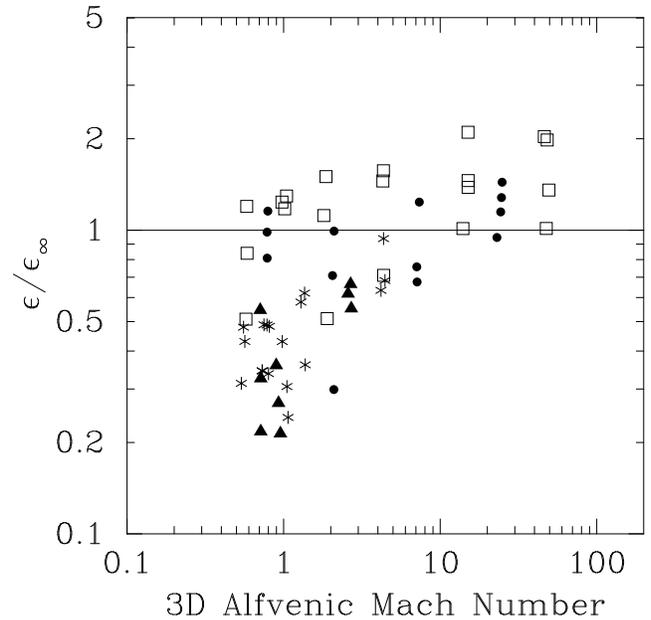}
\caption{Plot of $\epsilon$ (relative to the fitted $\epsilon_{\infty}({\mathcal{M}}_{1})$ relation in the hydrodynamic limit)
versus 3D Alfv\'{e}nic Mach number. Symbols denote the rms sonic Mach number in the simulation (open squares: ${\mathcal{M}}_{1} = 20$;
dots: ${\mathcal{M}}_{1} = 10$; triangles: ${\mathcal{M}}_{1} = 4$; asterixes: ${\mathcal{M}}_{1} = 2$).} 
\label{fig:evsma}
\end{figure}

Figure~\ref{fig:epicplot} shows representative fits to data taken from the latest snapshot of the grid simulations for 
both solenoidal and compressive forcing  (the other simulations yield similar results). It is clear that the proposed
form $h(\xi) = \xi^{-\epsilon}$ results in a good representation of the moment spectrum $\langle \xi^{q} v^{2}\rangle (q)$.
It is also apparent from the plots in Figure~\ref{fig:epicplot} that $\epsilon$ 
(i.e. the horizontal shift of the $\langle \xi^{q} v^{2}\rangle$ 
moment spectrum relative to the $\langle \xi^{q} \rangle$ moment spectrum) decreases with increasing rms Mach number.
To quantify this further, the fitted values of $\epsilon$ for all hydrodynamic data are shown 
in Figure~\ref{fig:epichydro}, plotted versus the measured 3D density-weighted Mach number, 
${\mathcal{M}}_{1} = \sigma_{1}/c_{s}$,
where $c_{s}$ is the sound speed. 
A power law relation is seen, represented by the fitted line, which is:
\begin{equation}
\epsilon_{\infty} = (0.38{\pm}0.05) {\mathcal{M}}_{1}^{-0.72{\pm}0.05}  ,  
\label{epsiloninfinity}
\end{equation}
where we have used the subscript $\infty$ to denote that the fit was obtained in the hydrodynamic
limit where the Alfv\'{e}nic Mach number ${\mathcal{M}}_{A} \rightarrow \infty$.

The values of $\epsilon$ obtained from the MHD simulations are shown in Figure~\ref{fig:epicmhd}, where
they are plotted versus ${\mathcal{M}}_{1}$. Relative to the hydrodynamic results, represented
by the straight line ($\epsilon_{\infty}$), the MHD fields tend to give lower values 
of $\epsilon$ at the lower Mach numbers. 
We find that in the low ${\mathcal{M}_{1}}$ regime, $\epsilon$ appears to decrease 
with decreasing Alfv\'{e}nic Mach Number, ${\mathcal{M}}_{A}$, as shown in Figure~\ref{fig:evsma}, where
we have plotted $\epsilon/\epsilon_{\infty}$ versus ${\mathcal{M}}_{A}$.
However, an ${\mathcal{M}}_{A}$-dependency (if any) is less clear at higher ${\mathcal{M}}_{1}$. 
In the analysis below, we will adopt $\epsilon_{\infty}$, obtained in the hydrodynamic limit,
as the basis for deriving correction factors ($g_{21}$). 

For a given $\langle \xi^{2} \rangle$, a smaller $\epsilon$ leads to a correction factor ($g_{21}$)
closer to unity. In fact, at fixed sonic Mach number, numerical simulations suggest that 
$\langle \xi^{2} \rangle$ should be closer to unity for strongly magnetized turbulence 
than for hydrodynamic turbulence (Molina et al 2012). Together with lower $\epsilon$, this 
would predict that $g_{21}$ should be closer to unity for
strongly magnetized turbulence than for the hydrodynamic case. 

Finally, as mentioned previously, we have only considered supersonic fields
in the above analysis. Since the velocity dispersion ratios (equation~(\ref{eq:vdisprat})) depend on
$\langle \xi^{q} \rangle$ to the power $-(m-n)\epsilon$, fields for which $\langle \xi^{q} \rangle \approx 1$ (i.e. uniform
or weakly varying density fields) require large values of $\epsilon$ for small corrections. We find that
measurement of $\epsilon$ in this regime (by the method given above) is rather unstable, with $\epsilon$ increasing
strongly, with large scatter, as $\langle \xi^{q} \rangle \longrightarrow 1$. In the analysis in Section~4, 
we will nevertheless investigate 
values of the correction factor $g_{21}$ extrapolated into the subsonic regime.

\subsection{Constraints from Observational Data}

In the preceding analysis, we made the assumption that the velocity dispersion decreases with the density at which it is measured (i.e.
equations~(\ref{vdispeq1})~and~(\ref{vdispeq2})), and explored the consequences on velocity dispersions calculated with a $\rho^{q}$--weight. 
To do this, priveleged access to the density and velocity fields in 3D is required, so this is not possible to do observationally
by exactly the same method. 
An alternative means of constraining the effect of such density--velocity correlations is to 
interpret equations~(\ref{vdispeq1})~and~(\ref{vdispeq2})
literally and examine velocity dispersions measured observationally in different density regimes. 

\subsubsection{Constraints from Larson's Relations}

Larson's (1981) relations between velocity dispersion (or linewidth) and cloud size ($\sigma_{v} \propto L^{a}$) and
between density and cloud size ($\rho \propto L^{-b}$) can provide a very crude measure of $\epsilon$ in molecular clouds. 
Combining the two relations gives a measure of $\epsilon \approx a/b$, though with some major caveats. The original
$a = 0.38$ and $b = 1.1$ derived by Larson (1981) give $\epsilon = 0.35$, while values of $a \approx 0.5$ and
$b = 1$ (assuming Virial equilibrium) give $\epsilon \approx 0.5$ from Solomon et al (1987). Large--scale CO surveys 
use size, linewidth, and mean density measured in distinct clouds rather than probing the density--dependence of velocity
dispersion internal to individual clouds. Extending such cloud/clump--based analyses to ``sub--cloud'' scales are highly questionable
(e.g. Ballesteros--Paredes \& Mac Low 2002; Schneider \& Brooks 2004), but nevertheless tend to yield values of
$a$ and $b$ roughly in accord with the large--scale values, though in the presence of significant scatter. 
Using the analysis of Simon et al (2001) on 4 inner Galaxy clouds (in Solomon et al 1987's survey region), we find 
values of $\epsilon$ between 0.17 and 0.39 (the lower values being mostly due to shallower linewidth--size relations than
that found by Solomon et al 1987).

Criticisms have been levelled at the density--size relation as being a consequence of limited dynamic range in cloud 
surface density (e.g. V\'{a}zquez-Semadeni, Ballesteros-Paredes, \& Rodriguez 1997). 
Using the Solomon et al (1987) cloud sample, Heyer et al (2009) found that, over the limited 
surface density range available, velocity dispersions rose with surface density as $\sigma_{v} \propto \Sigma^{1/2}$
at fixed cloud size. If we make the reasonable assumption that higher surface density indicates higher volume density,
then $a/b$ should provide an upper limit to $\epsilon$. 

The $\epsilon$ values derived here by Larson's relations are quoted for reference, and should be compared to
the better--motivated (and notably lower) values derived below, using density--selective tracers. 

\subsubsection{Constraints from Density--Selective Tracers}

A better--motivated idea than using Larson's relations is to compare velocity dispersions of trace molecules 
that are excited in different density regimes in the same cloud. We expect that high--density
tracers should have smaller velocity dispersions than low--density tracers, if their spectral lines are averaged over the same (large) volume. 
A rough estimate of $\epsilon$ may be arrived at by assuming that spectral line emission from a given trace molecule is dominated by contributions
from material near the molecule's critical density. With this assumption, if tracer $A$ has critical density $n_{c,A}$ and velocity 
dispersion $\sigma^{2}_{v,A}$ and tracer $B$ has critical density $n_{c,B}$ and velocity dispersion $\sigma^{2}_{v,B}$, then
equations~(\ref{vdispeq1})~and~(\ref{vdispeq2}) predict that:
\begin{equation}
\frac{\sigma^{2}_{v,A}}{\sigma^{2}_{v,B}} \approx \left(\frac{n_{c,A}}{n_{c,B}}\right)^{-\epsilon} .
\label{obsepsilon}
\end{equation}
From this, $\epsilon$ may be estimated via:
\begin{equation}
\epsilon \approx - \log \left(\frac{\sigma^{2}_{v,A}}{\sigma^{2}_{v,B}}\right) / \log \left(\frac{n_{c,A}}{n_{c,B}}\right) .
\label{obsepsilon2}
\end{equation}

\begin{figure*}
\includegraphics[width=115mm, angle=-90]{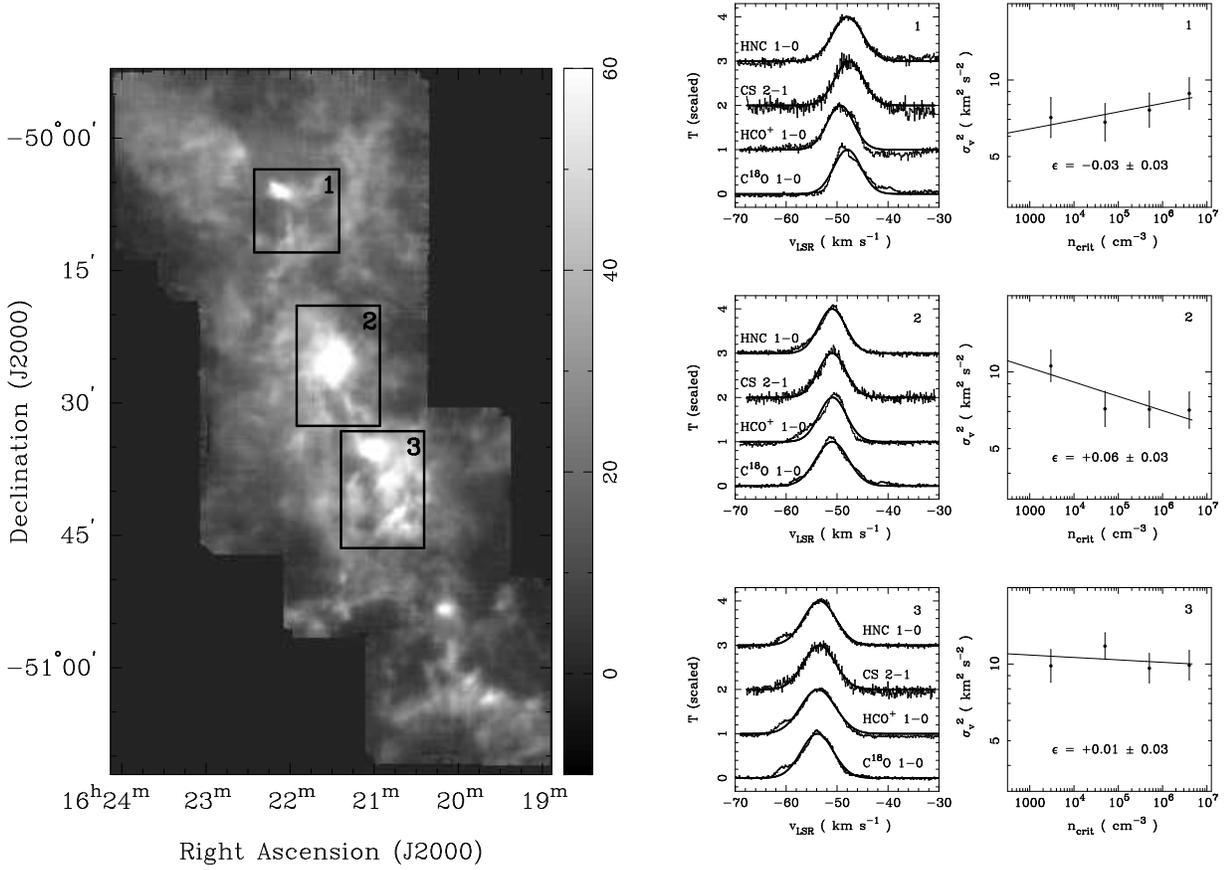}
\caption{Three sub-regions in the Delta Quadrant survey, in which the intensities of
C$^{18}$O 1--0, CS 2--1, HCO$^{+}$ 1--0, and HNC 1--0 spectral lines have been averaged and
fitted with gaussians to determine their velocity dispersions -- shown in the centre panel, with
spectra scaled to their peak and offset for clarity. In the right panel, we plot
log velocity dispersion versus log critical density to determine $\epsilon$ as
the slope of a fitted line for each region.}
\label{fig:g333spectra}
\end{figure*}

In reality, the tracers will sample a range of densities above their effective critical densities. A better motivation for
equations~(\ref{obsepsilon})~and~(\ref{obsepsilon2}) can be arrived at by more carefully considering such a system. If we
restrict the analysis to two tracers whose effective critical densities lie on the positive tail of the density PDF and assume
that in this regime, self-gravity will push the PDF into a power--law form ($P_{\xi}(\xi)$~$\propto$~$\xi^{-\alpha}$), then 
we can calculate velocity dispersions via equation~(\ref{eq:sigvxa}) as:
\begin{equation}
\sigma^{2}_{v,A} = \frac{\displaystyle\int_{\xi_{c,A}}^{\infty} {\mathrm{d}}\xi \; \xi^{-\alpha} \; \xi^{1-\epsilon} \sigma^{2}_{00} }{\displaystyle\int_{\xi_{c,A}}^{\infty} {\mathrm{d}}\xi \; \xi^{-\alpha} \; \xi}  = \left(\frac{2 - \alpha}{2 - \epsilon - \alpha}\right) \xi_{c,A}^{- \epsilon} \sigma^{2}_{00} , 
\label{powerlawtail}
\end{equation}
(and similarly for tracer $B$) and where we have taken the lower limit of the integral as $\xi_{c,A} = n_{c,A}/n_{0} = \rho_{c,A}/\rho_{0}$ and
assumed that $\alpha > q + 1 = 2$ so that the integrals converge. Note
that the velocity dispersion is calculated with a $\rho$--weight ($q = 1$).

Taking the velocity dispersion ratio of the two tracers, we find:
\begin{equation}
\frac{\sigma^{2}_{v,A}}{\sigma^{2}_{v,B}} = \left(\frac{\xi_{c,A}}{\xi_{c,B}}\right)^{- \epsilon}  ,
\label{powerlawtail2}
\end{equation}
which is exactly the same as equation~(\ref{obsepsilon}). If not a power--law PDF, a sufficiently
steeply--falling PDF (lognormal, exponential) will give similar results, as the dispersions are heavily
influenced by densities at the lower threshold (see e.g. Ballesteros--Paredes, D'Alessio, \& Hartmann 2012 for examination of a mathematically--equivalent system).

\subsubsection{Observational estimates of $\epsilon$}

In this Section, we will estimate a few contrasting values of $\epsilon$ from observational data.
 
McQuinn et al (2002) examined CS (J=2--1) and $^{13}$CO (J=1--0) emission in the inner Galaxy
observed as part of the Galatic Ring Survey (Jackson et al 2006) and found no significant
difference between CS (J=2--1) and $^{13}$CO (J=1--0) velocity dispersions, as evidenced by roughly constant
brightness temperature ratios for composite spectra averaged over large clouds (many parsecs scale). The critical densities
of CS (J=2--1) and $^{13}$CO (J=1--0) are $\sim$~10$^{3}$~cm$^{-3}$ and $\sim$~5~$\times$~10$^{5}$~cm$^{-3}$
respectively. Equation~(\ref{obsepsilon2}) in this case finds $\epsilon$ is ``very small''  (the
$g_{21}$ correction factor in our lognormal model above would therefore simply be $\sim$~unity -- i.e. no correction).
However, McQuinn et al (2002) ultimately concluded that subthermal excitation was a probable factor in the
line excitation (especially for CS) so that the effective critical densities would be lower than nominal.

To examine this more closely, we have used multi-tracer spectral line data, spanning a larger range in
critical density, from the Delta Quadrant Survey (see Lo et al 2009). 
Figure~\ref{fig:g333spectra} shows three regions
within the survey where spectral lines are single-component and allow easy fitting of gaussian functions
to estimate their dispersions. We have averaged the intensities over the boxes shown and determined
velocity dispersions for each of 4 transitions (C$^{18}$O 1--0, CS 2--1, HCO$^{+}$ 1--0, and HNC 1--0).
The spectra, with fitted gaussians, are also shown in Figure~\ref{fig:g333spectra}, along with the
variation of velocity dispersion with critical density of the tracer. From this, we
determine $\epsilon \approx 0$.

Williams \& Blitz (1998) examined CS (J=2--1) and $^{13}$CO (J=1--0) and (J=3--2) emission from a 
star--forming cloud (the Rosette nebula) and a non--star--forming cloud (G216, ``Maddalena's Cloud''). 
They found small differences in clump internal velocity dispersion in the (J=3--2) and (J=1--0) $^{13}$CO
lines: slightly broader 3--2 lines in the Rosette, relative to 1--0 (implying $\epsilon < 0$ in this case, since
the critical density of the 3--2 line is $\sim$~10 times that of the 1--0 line); 
slightly narrower lines are found for 3--2 in G216, relative to 1--0. They attribute this behaviour to local feedback
effects close to star--forming sources in the Rosette, which have no counterparts in the less active G216. Williams \& Blitz (1998)
do not report values for the linewidth ratios, but they can be estimated from their Figure~19; we will take 
$\sigma^{2}_{v,3-2}/\sigma^{2}_{v,1-0}$~$\approx$~2 and $\sigma^{2}_{v,3-2}/\sigma^{2}_{v,1-0}$~$\approx$~0.5
as representative for the Rosette and G216 respectively. These lead, assuming $n_{c,3-2}/n_{c,1-0}$~$\approx$~10,
to $\epsilon_{{\mathrm{Rosette}}}$~$\approx$~--0.3 and $\epsilon_{{\mathrm{G216}}}$~$\approx$~+0.3. These
values of $\epsilon$ are notably larger in magnitude than those found for our numerical simulations (and in the case of
the Rosette involve a sign change) but may not be representative values for {\it globally-determined} $\epsilon$ as
we require.  A negative value of $\epsilon$ probably cannot be
maintained over all densities (over all space), though one could potentially imagine high--density clumps moving through
a relatively static low density substrate as a possible configuration for this. The negative value of $\epsilon$ for the 
Rosette more likely comes instead from (e.g.) outflows injecting energy locally in a characteristic density regime 
near the J=3--2 critical density. Such behaviour could cause problems for our simple power law characterisation in
equation~(\ref{vdispeq2}).

McQuinn et al (2002) made large--scale averages of line profiles to produce their CS-$^{13}$CO comparisons, which is
more closely matched to our requirements. In contrast, Williams \& Blitz (1998)
examined linewidths from targeted clumps, where the role of density--velocity correlations are likely to be 
most emphasised. With this proviso in mind, we now examine the CS/$^{13}$CO (J=1--0) velocity dispersion ratios found
by Williams \& Blitz (1998). These are $\sigma^{2}_{v,{\mathrm{CS}}}/\sigma^{2}_{v,{\mathrm{13CO}}}$~$\sim$~0.7 (Rosette) and 
$\sigma^{2}_{v,{\mathrm{CS}}}/\sigma^{2}_{v,{\mathrm{13CO}}}$~$\sim$~0.5 (G216). Taking a critical density ratio of 500, we then
find $\epsilon_{{\mathrm{Rosette}}}$~$\approx$~+0.06 and $\epsilon_{{\mathrm{G216}}}$~$\approx$~+0.11. With a larger
baseline in critical density (and comparable $h\nu/k$ values for the transitions) these CS/$^{13}$CO--derived
values of $\epsilon$ are a more reliable measure than the 3--2/1--0--derived values, and are more in line with
our numerically-derived $\epsilon$--values.

We can estimate values of $\epsilon$ using C$^{18}$O (J=1--0) and N$_{2}$H$^{+}$ (J=1--0) data from the 
Perseus molecular cloud reported by Kirk et al (2010), who averaged spectra over spatially--extended
regions. The relative linewidths of C$^{18}$O and N$_{2}$H$^{+}$ vary amongst
the targeted regions (see their Figures 7--12). We estimate that 
$\sigma^{2}_{v,{\mathrm{N2H+}}}/\sigma^{2}_{v,{\mathrm{C18O}}}$ varies between $\sim$~0.2 and $\sim$~1 from
their graphs, and  will take an approximate value of 
$\sigma^{2}_{v,{\mathrm{N2H+}}}/\sigma^{2}_{v,{\mathrm{C18O}}}$~$\approx$~0.5 for the ensemble. 
Assuming the ratio of critical densities is $\sim$~1000, we find $\epsilon$~$\sim$~+0.1.

Walsh, Myers, \& Burton (2004) reported $^{13}$CO, C$^{18}$O, and N$_{2}$H$^{+}$ (all 1--0) linewidths
from a sample of nearby cores. These are targeted, single-point spectra towards separated regions, but 
taking an average over all spectra, we find that 
$<\sigma^{2}_{v,{\mathrm{N2H+}}}/\sigma^{2}_{v,{\mathrm{C18O}}}>$~=~0.66$\pm$0.37
and $<\sigma^{2}_{v,{\mathrm{N2H+}}}/\sigma^{2}_{v,{\mathrm{13CO}}}>$~=~0.29$\pm$0.16.
From these we derive $\epsilon$ values of +0.06 and +0.18, assuming a critical density ratio of
1000 in both cases. Within the uncertainties the dispersion ratios can be reconciled, though it is
clear that $^{13}$CO linewidths are broader than C$^{18}$O in most circumstances. 
The effective critical density for $^{13}$CO and C$^{18}$O
is likely to be different, due to abundance differences (i.e. $^{13}$CO is more abundant in the
lower density regions of the cloud than C$^{18}$O). In a medium
where the density PDF rises sharply towards lower densities, an abundant molecule's emission is in principle
subject to some, potentially significant, contribution from subthermally--excited regions where radiative
trapping is important, thereby lowering the critical density from its nominal value.
However, plausible variations (up to an order of magnitude difference)
in their effective critical densities cannot equalize the different $\epsilon$ values as $\epsilon$ is
only logarithmically sensitive to the assumed critical density ratio. 
We propose that the difference in $\epsilon$ can potentially be explained by path--length differences due to 
abundance effects: a velocity dispersion ratio of 
$<\sigma^{2}_{v,{\mathrm{13CO}}}/\sigma^{2}_{v,{\mathrm{C18O}}}>$~$\approx$~0.66/.29~$\approx$~2.28
can be explained by a path--length ratio of 2.28 in a medium where $\sigma^{2}_{v} \propto L$.

The small, positive nature of $\epsilon$ in most of the estimates above is in line with our initial expectation set out
in Section 4.1. Two factors can contribute to this. First, the velocity dispersion over some spatial scale $L$ in high 
density may be physically lower than the velocity dispersion over the same spatial scale $L$ in low density gas. 
Second, in a turbulent medium, the velocity dispersion increases with
spatial scale, and the lower density structures are necessarily more spatially--extended than the higher density structures.
For molecules with the same (nominal) critical density, the more abundant molecule will have a lower effective critical 
density, and therefore be more spatially--extended and have a higher velocity dispersion. (In principle, the first
of these two factors may be reversed (i.e. higher velocity dispersion in denser gas at fixed spatial scale) and
still yield a positive $\epsilon$ as long as the second factor dominates.)

If a targeted, single--point measurement is made towards an atypical position (e.g. a core) within a larger medium, 
as was done for some of the observations reported above, this may not provide a reliable measure of the density--dependence of velocity 
dispersion in the medium as a whole. Instead, if a velocity
dispersion measurement is made using a high--density tracer averaged over sufficiently large scales, then it will sample 
many density enhancements that are spatially distributed, and therefore result in a larger overall velocity dispersion 
(more comparable to the extended medium seen by a lower--density tracer). 

Though the $^{13}$CO (J=3--2) and (J=1--0) results above raise some questions, the observationally--estimated values of 
$\epsilon$ over large critical density baselines are in reasonable accord with those found in our numerical simulations.
We will take $0.05 \lesssim \epsilon \lesssim 0.3$ as defining the probable range of $\epsilon$ from the above calculations,
with values less than $\sim$~0.1 being favoured (i.e. as derived in cases where large--scale spatial averaging is conducted, 
and when the critical density span is larger).
The 3D $\rho$--weighted Mach numbers in the Rosette, G216, and Perseus are $\sim$~10--20, judging by the $^{13}$CO (J=1--0) 
linewidths, for kinetic temperatures of
$\sim$~10--20~K. The observational estimates compare reasonably well with the numerical results ($\epsilon$ versus ${\mathcal{M}}_{1}$) 
in Figure~\ref{fig:epichydro}.


\label{lastpage}

\end{document}